\documentclass[10pt,a4paper]{amsart}
\usepackage{amsfonts}
\usepackage{amsmath}
\usepackage{amssymb}
\usepackage{amsthm}
\usepackage{braket}
\usepackage{cite}
\usepackage{color}
\usepackage{geometry}
\usepackage{graphicx}
\usepackage{hyperref}
\usepackage{mathrsfs}
\usepackage{mathtools}
\usepackage{setspace}
\usepackage{stmaryrd}
\usepackage{tikz}

\newgeometry{top=3cm,bottom=3cm,outer=2cm,inner=2cm}

\numberwithin{equation}{section}

\newcommand{\bse}{\begin{subequations}}
\newcommand{\ese}{\end{subequations}}

\DeclareMathOperator{\tr}{Tr}
\DeclareMathOperator{\diag}{Diag}
\def\undertilde#1{\mathord{\vtop{\ialign{##\crcr
$\hfil\displaystyle{#1}\hfil$\crcr\noalign{\kern1.5pt\nointerlineskip}
$\hfil\widetilde{}\hfil$\crcr\noalign{\kern-6.5pt}}}}}
\def\underhat#1{\mathord{\vtop{\ialign{##\crcr
$\hfil\displaystyle{#1}\hfil$\crcr\noalign{\kern1.5pt\nointerlineskip}
$\hfil\widehat{}\hfil$\crcr\noalign{\kern-6.5pt}}}}}
\def\underbar#1{\mathord{\vtop{\ialign{##\crcr
$\hfil\displaystyle{#1}\hfil$\crcr\noalign{\kern1.5pt\nointerlineskip}
$\hfil\bar{}\hfil$\crcr\noalign{\kern-6.5pt}}}}}

\newcommand{\rd}{\mathrm{d}}

\newcommand{\bLd}{\mathbf{\Lambda}}
\newcommand{\tbLd}{{}^{t\!}\mathbf{\Lambda}}

\newcommand{\bo}{\mathbf{o}}
\newcommand{\tbo}{{}^{t\!}\mathbf{o}}
\newcommand{\bO}{\mathbf{O}}
\newcommand{\bOa}{\mathbf{\Omega}}

\newcommand{\bC}{\mathbf{C}}
\newcommand{\tbC}{{}^{t\!}\mathbf{C}}
\newcommand{\bU}{\mathbf{U}}
\newcommand{\tbU}{{}^{t\!}\mathbf{U}}
\newcommand{\bu}{\mathbf{u}}
\newcommand{\tbu}{{}^{t\!}\mathbf{u}}

\newcommand{\bc}{\mathbf{c}}
\newcommand{\tbc}{{}^{t\!}\mathbf{c}}
\newcommand{\bA}{\mathbf{A}}

\newcommand{\bW}{\mathbf{W}}
\newcommand{\bM}{\mathbf{M}}

\newcommand{\bV}{\mathbf{V}}

\newcommand{\ld}{\lambda}

\newcommand{\oa}{\omega}
\newcommand{\Oa}{\Omega}
\newcommand{\bx}{\mathbf{x}}
\newcommand{\br}{\mathbf{r}}
\newcommand{\bs}{\mathbf{s}}
\newcommand{\bK}{\mathbf{K}}

\title[Linear integral equations, infinite matrices, and soliton hierarchies]{Linear integral equations, infinite matrices, \\ and soliton hierarchies}
\author{Wei FU}
\author{Frank W Nijhoff}
\address{School of Mathematics, University of Leeds, Leeds LS2 9JT, UK}

\begin{document}

\maketitle

\begin{abstract}
A systematic framework is presented for the construction of hierarchies of soliton equations. This is realised by considering scalar linear integral 
equations and their representations in terms of infinite matrices, which give rise to all (2+1)- and (1+1)-dimensional soliton hierarchies associated 
with scalar differential spectral problems. The integrability characteristics for the obtained soliton hierarchies, including Miura-type transforms, 
$\tau$-functions, Lax pairs as well as soliton solutions, are also derived within this framework. 
\end{abstract}

\section{Introduction}\label{S:Intro}

Integrable nonlinear partial differential equations (PDEs) arise in a variety of areas in modern mathematics and physics. 
Over the past few decades, there have been many methods developed for the construction and solutions for such equations, including inverse scattering 
method, Riemann--Hilbert approach, finite-gap integration, Hirota's bilinear method and methods based on representation theory, cf. the monographs 
\cite{AC91,NMPZ84,Hir04,Kac94}. 
Among those models, the Kadomtsev--Petviashvili (KP) hierarchy is often considered to be the most fundamental one and the most popular construction of 
this hierarchy is Sato's approach based on pseudo-differential operators \cite{SM81}.  This is related to the observation that the KP hierarchy 
is closely related to infinite-dimensional Grassmannians \cite{Sat81}. 
This idea was further developed by Date, Jimbo, Kashiwara and Miwa who classified soliton equations by using transformation groups associated 
with infinite-dimensional Lie algebras based on the so-called fundamental bilinear identity leading to the hierarchy of equations in Hirota's 
form in terms of the $\tau$-function (cf. e.g. \cite{JM83,MJD00} and references therein for the original research papers by the Kyoto school).

The pseudo-differential operator approach may have some disadvantages. One disadvantage is that it singles out a preferred independent variable from all 
the flow variables in the hierarchy whereas the latter can be considered all to be on the same footing. The other disadvantage is that it can only be 
discretised to obtain semi-discrete equations but no fully-discrete equations have been obtained yet from this approach. In contrast, in this paper we will 
promote the direct linearisation (DL) approach for treating the KP hierarchy in which no preselected independent variable is required to set up the 
constitutive relations, and which also allows in a natural way to a full discretisation. 

The DL was first proposed by Fokas and Ablowitz in \cite{FA81} for solving the Korteweg--de Vries (KdV) and Painlev\'e II equation, and later extended 
to the KP equation in \cite{FA83} and some other nonlinear equations in \cite{AFA83,SAF84}). The idea behind the DL approach, to use a general linear 
singular integral equations in the spectral variable as the starting point, was developed further in a series of papers, cf. e.g. 
\cite{NQC83,QNCL84,NLQCV82,Nij88}. As a powerful tool to treat these linear integral equations, an infinite matrix structure was introduced, cf. 
\cite{NLQCV82,Nij88,FN16}, which allows to capture whole Miura-related families of equations together with their hierarchies, within one framework. 
Furthermore, in this framework, all the independent variables (discrete as well as continuous) are treated on the same footing. 

In a recent paper \cite{FN16} the DL was established for the three families of the discrete KP-type equations, namely the discrete AKP, BKP and CKP equations, 
extending the earlier results on discrete KP equations of A-type \cite{NCWQ84}. Based on the new insights provided by the previous paper, in the current 
paper we revisit the continuous hierarchies for the AKP, BKP and CKP equations as well as their dimensional reductions. The resulting (1+1)-dimensional 
hierarchies include the following examples: the KdV, Boussinesq (BSQ), generalised Hirota--Satsuma (gHS), Hirota--Satsuma (HS), Sawada--Kotera (SK) and 
Kaup--Kupershmidt (KK), bidirectional SK (bSK) and bidirectional KK (bKK), and Ito hierarchies. The DL framework is presented in the language of infinite 
matrices and this treatment provides the following: 
 {\it i)} different nonlinear forms in the each class together with the Miura-type transforms; 
 {\it ii)} the multilinear form in terms of the $\tau$-function for each class; 
 {\it iii)} the Lax pairs of both nonlinear equations and the multilinear equation for each class; 
 {\it iv)} the soliton solutions for both nonlinear and multilinear equations. 
In spite of the undeniable virtues of other approaches, we believe that the DL framework provides the most comprehensive treatment of all these results.

The paper is organised as follows: In Section \ref{S:DL} we set up the ingredients of the infinite matrix structure and explain the DL scheme in this language. 
Section \ref{S:3D} is contributed to the three important (2+1)-dimensional soliton hierarchies, namely, the AKP, BKP and CKP hierarchies. The dimensional 
reductions of the higher-dimensional models are discussed in Section \ref{S:2D} which includes a number of (1+1)-dimensional integrable models. Finally in 
Section \ref{S:Solution}, soliton solutions for all the soliton hierarchies will be given as particular cases as the general DL framework. 

\section{Infinite matrices and linear integral equations}\label{S:DL}

\subsection{Infinite matrices and vectors}

The details of infinite matrices can be found in \cite{FN16}. In this subsection we only give a brief introduction to the notion which include all the 
properties and operations we need in the paper.

We first of all introduce three fundamental objects $\bLd$, $\tbLd$ and $\bO$ and they obey the relation
\begin{align}\label{InfMatRule}
 \bO\cdot\tbLd^i\cdot\bLd^j\cdot\bO=\delta_{-i,-j}\bO, \quad \forall i,j\in\mathbb{Z},
\end{align}
where $\tbLd^i$ and $\bLd^j$ mean the $i$th and $j$th compositions of $\tbLd$ and $\bLd$ respectively. In particular $\tbLd^0$ and $\bLd^0$ are defined 
as the identities and the relation in this case reads $\bO^2=\bO$, namely, the object $\bO$ is a projector. In general, $\bLd$, $\tbLd$ and $\bO$ do not 
commute with each other but they commute with the elements from a suitably chosen field of functions $\mathcal{F}$ which will be determined later. 
We consider infinite matrices of the form:  
$\bU$ is defined as 
\begin{align}\label{InfMat}
 \bU=\sum_{i,j\in\mathbb{Z}}U_{i,j}\bLd^{-i}\cdot\bO\cdot\tbLd^{-j},
\end{align}
where $U_{i,j}$ are elements in the field $\mathcal{F}$. Here $U_{i,j}$ can be understood as the entries in the infinite matrix, while  
$\bLd^{-i}\cdot\bO\cdot\tbLd^{-j}$ with $i,j\in\mathbb{Z}$ can be thought of as a basis in this infinite matrix space. In fact, it can 
be verified that the infinite matrices defined in this way constitute a linear space where the addition and the scalar multiplication 
are the same as those in the finite case. The multiplication of two infinite matrices is also well defined if one follows the rule 
\eqref{InfMatRule} that the objects must obey. The transposition of $\bU$ is defined as 
$\tbU=\sum_{i,j}U_{i,j}\bLd^{-j}\cdot\bO\cdot\tbLd^{-i}$. If $\bU=\tbU$, comparing the coefficients (i.e. the matrix entries) of them, 
we immediately have $U_{i,j}=U_{j,i}$. Likewise, we have $U_{i,j}=-U_{j,i}$ if $\bU=-\tbU$. These results coincide with the properties in 
finite matrices. Furthermore, one can introduce the action $(\,\cdot\,)_{0,0}$ on an arbitrary infinite matrix defined by taking the coefficient of 
$\bLd^0\cdot\bO\cdot\tbLd^0$ in the definition of an infinite matrix. Taking $\bU$ defined in \eqref{InfMat} as an example, we have that
\begin{align}\label{Lambda}
 (\bLd^{i_0}\cdot\bU\cdot\tbLd^{j_0})_{0,0}=U_{i_0,j_0}, \quad \forall i_0,j_0\in\mathbb{Z}.
\end{align}

As the notion of infinite matrices have been defined, we now define infinite vectors. Suppose $\bo$ and $\tbo$ are two vectors obeying the relations
\begin{align}\label{InfVecRule}
 \tbo\cdot\tbLd^i\cdot\bLd^j\cdot\bo=\delta_{-i,-j}, \quad \bo\cdot\tbo=\bO, \quad i,j\in\mathbb{Z}.
\end{align}
An infinite vector $\bu$ and its adjoint vector $\tbu$ are defined as
\begin{align}\label{InfVec}
 \bu=\sum_{i\in\mathbb{Z}}u^{(i)}\bLd^{-i}\cdot\bo, \quad \tbu=\sum_{i\in\mathbb{Z}}u^{(i)}\tbo\cdot\tbLd^{-i},
\end{align}
where $u^{(i)}$ are elements in the same field $\mathcal{F}$. Following the rules in \eqref{InfVecRule}, it can easily be verified that the addition, 
the scalar multiplication of infinite vectors as well as the multiplication of an infinite vector and an adjoint vector are all well-defined, 
i.e. the addition of two infinite vectors is still an infinite vector, an infinite vector multiplied by a scalar quantity is still an infinite vector,
and the multiplication of an infinite vector and an adjoint vector is an infinite matrix while the multiplication of an adjoint vector and an infinite 
vector is a scalar quantity. It is similar to the case in infinite matrices that we can also define the action $(\,\cdot\,)_0$ as taking the coefficient 
of $\bLd^0\cdot\bo$ for an arbitrary infinite vector $\bu$ and that of $\tbo\cdot\tbLd^0$ for an arbitrary adjoint infinite vector $\tbu$ respectively. 
For instance, one can deduce that 
\begin{align*}
 (\bLd^{i_0}\cdot\bu)_0=u^{(i_0)}, \quad (\tbu\cdot\tbLd^{j_0})_0=u^{(j_0)}, \quad \forall i_0,j_0\in\mathbb{Z}.
\end{align*}
The multiplication of an infinite matrix and an infinite vector is also well-defined if one follows from \eqref{InfMatRule} and \eqref{InfVecRule}. 
It obeys exactly the same rules as those in the theory of finite matrices and vectors. In other words, $\bU\cdot\bu$ is an infinite vector and 
$\tbu\cdot\bU$ is an adjoint infinite vector. 

We will also need to use the notion of trace and determinant of an infinite matrix in the paper where the infinite matrix $\bO$ is always involved 
in concrete calculation. In fact, trace and determinant can be defined in the usual way and in the case of having the projector $\bO$, the issue of 
convergence of summations in the definitions no longer appears. We can easily verify that the trace obeys the property 
$\tr(\bU\cdot\bO)=\tr(\bO\cdot\bU)=(\bU)_{0,0}$. Besides, we also have $\tr(\bU+\bV)=\tr\bU+\tr\bV$ as well as $\tr(\bU\cdot\bV)=\tr(\bV\cdot\bU)$. 
As for the determinant of infinite matrices, we have 
\begin{align*}
 \det(1+\bU\cdot\bO\cdot\bV)=\det(1+\bU\cdot\bo\cdot\tbo\cdot\bV)=1+\tbo\cdot\bV\cdot\bU\cdot\bo=1+(\bV\cdot\bU)_{0,0},
\end{align*}
where $\bU\cdot\bo$ and $\tbo\cdot\bV$ are thought of as an infinite vector and an adjoint infinite vector respectively and \eqref{InfMatRule} 
and \eqref{InfVecRule} are used. This identity is an infinite version of the important formula 
$\det(1+\mathbf{a}\cdot\mathbf{b}^{\mathrm{T}})=1+\mathbf{b}^{\mathrm{T}}\cdot\mathbf{a}$ for finite column vectors $\mathbf{a},\mathbf{b}$, 
namely, the Weinstein--Aronszajn formula in rank 1 case, which evaluates the determinant of an infinite matrix in terms of a scalar quantity. 

\subsection{Infinite matrix representation of linear integral equations}

The starting point in the framework is a linear integral equation associated with (2+1)-dimensional integrable systems: 
\begin{align}\label{IntEq}
 \bu_k+\iint_{D}\rd\zeta(l,l')\rho_k\Omega_{k,l'}\sigma_{l'}\bu_l=\rho_k\bc_k,
\end{align}
where $\rd \zeta(l,l')$ is a certain measure for the double integral on a domain $D$ in the space of the spectral variables $l$ and $l'$. 
The wave function $\bu_k=\sum_{i\in\mathbb{Z}}u_k^{(i)}\bLd^{-i}\cdot\bo$ is an infinite vector where $u_k^{(i)}=u_k^{(i)}(\bx;k)$ with 
$\bx=\{x_j|j\in\mathbb{Z}^+\}$ are $C^{\infty}$-functions with respect to the independent variables in $\bx$ and $k$ is the spectral parameter. 
$\bc_k$ is also an infinite vector which is defined as $\bc_k=\sum_{i\in\mathbb{Z}}k^i\bLd^{-i}\cdot\bo$. While $\Omega_{k,l'}$ is the Cauchy kernel for 
the integral equation and $\rho_k$, $\sigma_{l'}$ are the plane wave factors which will be given later when we deal with concrete equations. One remark 
here is that the measure $\rd\zeta(l,l')$ depends on $l$ and $l'$ and we must have a double integral involved. Once the measure collapses the integral 
equation turns out to be a local Riemann--Hilbert problem leading to (1+1)-dimensional integrable systems and this will be done in Section \ref{S:2D} as 
dimensional reductions of higher-dimensional equations. Another remark is that actually we only need three variables to represent a (2+1)-dimensional 
integrable equation, but we still introduce an infinite number of variables in order to describe the whole hierarchy of a soliton equation. For instance, 
one can fix $x_1,x_2$ as the space variables and the flexible $x_j$ for $j\in\mathbb{Z}^+$ is the time variable which denotes the $j$th flows in the 
hierarchy. The third remark is that although $\bu_k$ is an infinite vector, nevertheless, the equation \eqref{IntEq} is a set of many integral equations 
in terms of the wave functions $u_k^{(i)}$. Therefore it should still be understood as a set of scalar integral equations. 

Now we introduce some key quantities in the representation of the linear integral equation in terms of infinite matrices. We define 
the infinite matrices 
\begin{align}\label{AKP:U}
 \bC=\iint_D\rd\zeta(l,l')\rho_l\bc_l\tbc_{l'}\sigma_{l'}, \quad \bU=\iint_D\rd\zeta(l,l')\bu_l\tbc_{l'}\sigma_{l'}.
\end{align}
In fact, $\bC$ can be understood as the infinite matrix analogue of the plane wave factor $\rho_l$ and $\sigma_{l'}$ and $\bU$ can be considered as 
the infinite matrix version for the wave function $\bu_k$. Next we define an operator $\bOa$ by $\Oa_{k,l'}=\tbc_{l'}\cdot\bOa\cdot\bc_k$ and it is clear 
to see that $\bOa$ is the analogue of the Cauchy kernel in our infinite matrix language. Consider the new quantities $\bC$, $\bU$ and $\bOa$, one can now 
represent the integral equation \eqref{IntEq} as 
\begin{align}\label{uk}
 \bu_k=(1-\bU\cdot\bOa)\cdot\rho_k\bc_k
\end{align}
and its nonlinear version, namely, the relation for the infinite matrix $\bU$, can be written in the form of 
\begin{align}\label{U}
 \bU=(1-\bU\cdot\bOa)\cdot\bC.
\end{align}
Equations \eqref{uk} and \eqref{U} are the fundamental relations in the DL scheme. Once $\bOa$ and $\bC$ are given, one can then consider the dynamical 
evolutions for both quantities. By choosing entries/components in $\bU$ and $\bu_k$ suitably, closed-form nonlinear equations and the associated linear 
problems will arise. The $\bOa$ and $\bC$ are also the ingredients for the $\tau$-function which can represent the multilinear structure for the nonlinear 
equations. Then the $\bU$ defined as a double integral in \eqref{AKP:U} provides the most general solution for the obtained closed-form equations. Solutions 
having a particular behaviour (e.g. soliton solutions) can be constructed by taking a suitable form of the measure $\rd\zeta(l,l')$ and the domain $D$. 
In the paper we only consider soliton solutions of integrable equations as they can be regarded as one of the most important characteristics of integrability.

\section{(2+1)-dimensional soliton hierarchies}\label{S:3D}

In this section, some particular $\bC$ and $\bOa$ are given and the DL scheme is established. The resulting models include the AKP, BKP and CKP 
hierarchies. The three hierarchies play the role of the master integrable systems in our framework and in Section \ref{S:2D} we will see that 
they generate a number of (1+1)-dimensional soliton hierarchies by suitable dimensional reductions.

\subsection{The AKP hierarchy}\label{S:AKP}
The AKP hierarchy, normally known as the KP hierarchy, is associated with the infinite-dimensional Lie algebra $A_{\infty}$, namely $\mathfrak{gl}(\infty)$. 
In this class, we consider a particular infinite matrix $\bC$ given by 
\begin{align}\label{AKP:C}
 \bC=\iint_D\rd\zeta(l,l')\rho_l\bc_l\tbc_{l'}\sigma_{l'}, \quad 
 \rho_k=\exp\Big(\sum_{j=1}^\infty k^jx_j\Big), \quad \sigma_{k'}=\exp\Big(-\sum_{j=1}^\infty (-k')^jx_j\Big),
\end{align}
where $\rho_k$ and $\sigma_{k'}$ are the plane wave factors. Differentiating $\bC$ with respect to $x_j$ and notice the form of 
$\rho_k$ and $\sigma_{k'}$, one can obtain the dynamical evolution of $\bC$: 
\begin{align}\label{AKP:CDyn}
 \partial_j\bC=\bLd^j\cdot\bC-\bC\cdot(-\tbLd)^j, \quad j\in\mathbb{Z}^+,
\end{align}
where $\partial_j\doteq\partial_{x_j}$. The operator $\bOa$, namely, the infinite matrix version of the Cauchy kernel in this case is defined by the relation
\begin{align}\label{AKP:Oa}
 \bOa\cdot\bLd+\tbLd\cdot\bOa=\bO.
\end{align}
In fact, the Cauchy kernel can be recovered from $\bOa$ and it is now given by $\Omega_{k,l'}=\frac{1}{k+l'}$. One can now consider the 
dynamical evolution of $\bU$ defined as \eqref{U}. In fact, making use of Equations \eqref{AKP:CDyn} and \eqref{AKP:Oa}, we end up with the relation 
for $\bU$ as follow: 
\begin{align}\label{AKP:UDyn}
 \partial_j\bU=\bLd^j\cdot\bU-\bU\cdot(-\tbLd)^j-\bU\cdot\bO_j\cdot\bU, \quad j\in\mathbb{Z}^+,
\end{align}
where $\bO_j\doteq\sum_{i=0}^{j-1}(-\tbLd)^i\cdot\bO\cdot\bLd^{j-1-i}$. Equations \eqref{AKP:UDyn} can be considered as the AKP hierarchy 
expressed by the infinite matrix $\bU$.

Following from the definition of $\bO_j$, one can easily prove the recurrence relation 
\begin{align}\label{ORecur}
 \bO_{i+j}=\bO_i\cdot\bLd^j+(-\tbLd)^i\cdot\bO_j.
\end{align}
Making use of the recurrence relation of $\bO_j$ \eqref{ORecur}, one can find the recurrence relation for the dynamical relation of $\bU$ 
\eqref{AKP:UDyn} via some straightforward computation and they are given by: 
\begin{align*}
 &(\partial_{i+j}+\partial_i\partial_j)\bU=(\bLd^i-\bU\cdot\bO_i)\cdot(\partial_j\bU)+(\bLd^j-\bU\cdot\bO_j)\cdot(\partial_i\bU), \\
 &(\partial_{i+j}-\partial_i\partial_j)\bU=(\partial_j\bU)\cdot((-\tbLd)^i+\bO_i\cdot\bU)+(\partial_i\bU)\cdot((-\tbLd)^j+\bO_j\cdot\bU).
\end{align*}
The importance of the above relations is that in the first one only $\bLd$ is involved and in the second one only $\tbLd$ is involved. 
This observation provides us with a possibility to reduce the degree of $\bLd$ and $\tbLd$ in the dynamical relation of $\bU$, 
i.e. \eqref{AKP:UDyn}. In fact, taking $i=j=1$, one can have 
\bse\label{AKP:ReducOrder}
\begin{align}
 &\bLd\cdot\bU=\frac{1}{2}\partial_1^{-1}(\partial_2\bU+\partial_1^2\bU+2\bU\cdot\bO\cdot(\partial_1\bU)), \\
 &\bU\cdot(-\tbLd)=\frac{1}{2}\partial_1^{-1}(\partial_2\bU-\partial_1^2\bU-2(\partial_1\bU)\cdot\bO\cdot\bU).
\end{align}
\ese
With the help of Equation \eqref{AKP:ReducOrder}, one can eliminate $\bLd$ and $\tbLd$ in \eqref{AKP:UDyn} and obtain a differential relation 
for only $\bU$ and $\bO$. Consider the entry $U_{0,0}=(\bU)_{0,0}\doteq u$ in the infinite matrix $\bU$, the KP (i.e. AKP) hierarchy can be derived. 
The first nontrivial equation (when $j=3$), namely, the KP equation, is given by 
\begin{align}\label{AKP:NL}
 u_{x_3x_1}=\Big(\frac{1}{4}u_{x_1x_1x_1}+\frac{3}{2}u_{x_1}^2\Big)_{x_1}+\frac{3}{4}u_{x_2x_2}.
\end{align}
We note that a slightly different derivation of the KP equation \eqref{AKP:NL} from the DL framework can be found in \cite{Wal01}. 
Other nonlinear forms in the AKP class can also be derived if one defines $v=[\ln(1-U_{0,-1})]_{x_1}$, $w=-U_{1,-1}/(1-U_{0,-1})$ and $z=U_{-1,-1}-x_1$. 
In fact, \eqref{AKP:UDyn} leads to some Miura-type transforms between these variables:
\begin{align}\label{AKP:MT}
 2u_{x_1}=-v_{x_1}-v^2+\partial_{x_1}^{-1}v_{x_2}, \quad 
 v=\frac{1}{2}\frac{z_{x_1x_1}+z_{x_2}}{z_{x_1}}.
\end{align}
We omit the derivation of these transforms and one can verify these identities by substituting the derivatives with the entries in the infinite matrix $\bU$ 
by using \eqref{AKP:UDyn}. By these transforms, one can from \eqref{AKP:NL} derive 
\begin{align}
 &v_{x_3x_1}=\Big(\frac{1}{4}v_{x_1x_1x_1}-\frac{3}{2}v^2v_{x_1}\Big)_{x_1}+\frac{3}{2}v_{x_1x_1}\partial_{x_1}^{-1}v_{x_2}+\frac{3}{2}v_{x_1}v_{x_2}
 +\frac{3}{4}v_{x_2x_2}, \label{AKP:Mod} \\
 &\Big(\frac{z_{x_3}}{z_{x_1}}\Big)_{x_1}=\frac{1}{4}\{z,x_1\}_{x_1} +\frac{3}{4}\frac{z_{x_2}}{z_{x_1}}\Big(\frac{z_{x_2}}{z_{x_1}}\Big)_{x_1}
 +\frac{3}{4}\Big(\frac{z_{x_2}}{z_{x_1}}\Big)_{x_2}, \quad 
 \{z,x_1\}\doteq \frac{z_{x_1x_1x_1}}{z_{x_1}}-\frac{3}{2}\frac{z_{x_1x_1}^2}{z_{x_1}^2}. \label{AKP:Sch}
\end{align}
The two equations are referred to as the modified KP (mKP) equation and the Schwarzian KP (SKP) equation respectively. $\{z,x_1\}$ defined above 
is the Schwarzian derivative of $z$ with respect to $x_1$ and it is M\"obius invariant, namely it is invariant under a fractionally linear transform 
and therefore it is clear to see that the SKP equation has a M\"obius symmetry (cf. \cite{Wei83}). In addition, one can also find 
$w_{x_1}=\tfrac{1}{2}(v_{x_1}-v^2+\partial_{x_1}^{-1}v_{x_2})$, which implies that the variable $w$ also obeys the KP equation 
\eqref{AKP:NL}. One remark here is that the AKP class can be obtained from different $\bOa$. For instance, one can replace $\bO$ in \eqref{AKP:Oa} by 
$\tfrac{1}{2}(\bO\cdot\bLd-\tbLd\cdot\bO)$, then $U_{0,0}$ will also give us a slightly different form (a weak form) of the mKP equation. 

From the above result we have seen that the AKP hierarchy has a lot of nonlinear forms. Therefore we need a quantity that can describe the evolution 
of the AKP hierarchy in a unified way. The $\tau$-function defined by $\tau=\det(1+\bOa\cdot\bC)$ can actually be a very good candidate as it contains 
the information of $\bOa$ and $\bC$ that is given at the beginning of our scheme. This reminds us of considering the dynamical evolution of the 
$\tau$-function. In fact some simple calculation using the rank 1 Weinstein--Aronszajn formula shows that 
\begin{align}\label{AKP:tau}
 \partial_j\ln\tau=\partial_j\ln(\det(1+\bOa\cdot\bC))=\tr(\bO_j\cdot\bU),
\end{align}
where in the derivation the identity $\ln(\det\bW)=\tr(\ln\bW$) for an arbitrary matrix $\bW$ is used. When $j=1$, it gives us the multilinear transform 
$u=(\ln\tau)_{x_1}$ and as a result the equation \eqref{AKP:NL} turns out to be the bilinear equation
\begin{align}\label{AKP:BL}
 (D_1^4-4D_1D_3+3D_2^2)\tau\cdot\tau=0,
\end{align}
in which the operator $D_j$ is the standard bilinear operator in terms of the continuous independent variable $x_j$ defined by
\begin{align*}
 D_j^nf(\bx)\cdot g(\bx')=(\partial_{x_j}-\partial_{x_j'})^nf(\bx)g(\bx')|_{\bx'=\bx}, \quad \bx=\{x_j|j\in\mathbb{Z}^+\}, \quad \bx'=\{x_j'|j\in\mathbb{Z}^+\}.
\end{align*}
The higher-order equations for the $\tau$-function can be obtained from the nonlinear AKP hierarchy in $u$ via the same transform. But we note that 
the obtained $\tau$-equations may no longer be bilinear, instead they will be multilinear. It is this set of multilinear equations that governs the 
algebraic structure behind the AKP class. 

In the DL framework, the fundamental object is the infinite matrix $\bU$ which generates the whole KP hierarchy 
with the help of \eqref{AKP:UDyn} and \eqref{AKP:ReducOrder}; more precisely, eliminating $\bLd$ and $\tbLd$ in these relations and 
considering the $(0,0)$-entry gives us the whole KP hierarchy expressed by the nonlinear variable $u$. The bilinear transform $u=(\ln\tau)_{x_1}$ 
then brings us a hierarchy of multilinear equations expressed by the same $\tau$-function. These multilinear equations are deep down equivalent to 
the bilinear KP equations in the Sato scheme, see e.g. \cite{JM83} and \cite{OHTI93}. The difference is that in the bilinear framework more 
and more independent variables must be involved in higher-order bilinear equations, but in our approach each multilinear equation only 
depends on three dynamical variables, namely $x_1$, $x_2$ and $x_j$. 

Now we consider the linear problem of the AKP hierarchy. The differentiation of \eqref{uk} together with Equations \eqref{AKP:UDyn} and \eqref{AKP:Oa} bring 
us the dynamical evolution for $\bu_k$ in the AKP class given by 
\begin{align}\label{AKP:ukDyn}
 \partial_j\bu_k=\bLd^j\cdot\bu_k-\bU\cdot\bO_j\cdot\bu_k.
\end{align}
Like how we deal with the nonlinear variable $\bU$, a similar derivation provides us with the following important relations for $\bu_k$:
\begin{align*}
 (\partial_{i+j}-\partial_i\partial_j)\bu_k=(\partial_j\bU)\cdot\bO_i\cdot\bu_k+(\partial_i\bU)\cdot\bO_j\cdot\bu_k, \quad j\in\mathbb{Z}^+.
\end{align*}
By taking $i=j=1$ and setting $\phi=u_k^{(0)}$ in the above relation, the linear problem can be obtained and it is as the following: 
\bse\label{AKP:Lax}
\begin{align}\label{AKP:LaxS}
 \phi_{x_2}=(\partial_1^2+2u_{x_1})\phi.
\end{align}
The linear equation \eqref{AKP:LaxS} in $\phi$ governs the linear structure of the whole AKP hierarchy, namely, it is the spatial part of 
the Lax pairs for all the members in the AKP hierarchy. The temporal evolutions of the hierarchy can be derived by considering the corresponding 
flows separately. In practice, one can have from 
\eqref{AKP:ukDyn} 
\begin{align*}
\bLd\cdot\bu_k=\partial_1\bu_k+\bU\cdot\bO\cdot\bu_k
\end{align*}
by taking $j=1$. This relation together with \eqref{AKP:ReducOrder} can help to reduce the order of $\bLd$ and $\tbLd$ in $\partial_j\bu_k$ 
following from \eqref{AKP:ukDyn} and therefore the temporal evolutions can be derived. For instance, if we consider the $x_3$-flow in \eqref{AKP:ukDyn}, 
we have the temporal evolution 
\begin{align}\label{AKP:LaxT}
 \phi_{x_3}=\Big[\partial_1^3+3u_{x_1}\partial_1+\frac{3}{2}(u_{x_1x_1}+u_{x_2})\Big]\phi.
\end{align}
\ese
The compatibility condition of the spatial part \eqref{AKP:LaxS} and the temporal evolution \eqref{AKP:LaxT}, namely, $\phi_{x_2x_3}=\phi_{x_3x_2}$, 
gives us the KP equation \eqref{AKP:NL}. The Lax pairs for the other nonlinear and multilinear forms can be calculated by replacing $u$ in \eqref{AKP:Lax} 
by the other variables via the Miura-type transforms \eqref{AKP:MT} and the multilinear transform. The temporal evolution of the higher-order equations 
in the hierarchy can be derived in a similar way.

\subsection{The BKP hierarchy}\label{S:BKP}
The BKP hierarchy is associated with the infinite-dimensional algebra $B_\infty$ which is a sub-algebra of $A_\infty$, therefore the BKP hierarchy can 
be understood as a sub-hierarchy of AKP. In this case, we take a particular infinite matrix $\bC$ as follow: 
\begin{align}\label{BKP:C}
 \bC=\iint_D\rd\zeta(l,l')\rho_l\bc_l\tbc_{l'}\rho_{l'}, \quad \rho_k=\exp\Big(\sum_{j=0}^\infty k^{2j+1}x_{2j+1}\Big), \quad 
 \rd\zeta(l',l)=-\rd\zeta(l,l').
\end{align}
From the above formula one can see the difference between BKP and AKP is that in the BKP hierarchy only odd independent variables are involved and besides 
the measure in the integral has the antisymmetry property. One can easily find that $\bC$ satisfies the dynamical relation 
\begin{align}\label{BKP:CDyn}
 \partial_{2j+1}\bC=\bLd^{2j+1}\cdot\bC+\bC\cdot\tbLd^{2j+1}, \quad j=0,1,2,\cdots
\end{align}
as well as the antisymmetry property $\tbC=-\bC$. Now we require that the infinite matrix version $\bOa$ of the kernel obeys the algebraic relation 
\begin{align}\label{BKP:Oa}
 \bOa\cdot\bLd+\tbLd\cdot\bOa=\frac{1}{2}(\bO\cdot\bLd-\tbLd\cdot\bO).
\end{align}
Actually we have pointed out in the previous section that replacing the right hand side of \eqref{AKP:Oa} by $\tfrac{1}{2}(\bO\cdot\bLd-\tbLd\cdot\bO)$ 
also gives us the AKP hierarchy in a slightly different form. So Equation \eqref{BKP:Oa} is just another representation of \eqref{AKP:Oa}. 
In other words, the infinite matrix relation for $\bOa$ is preserved from AKP to BKP, but here we impose in addition the antisymmetry property on the 
measure as one can see in \eqref{BKP:C}. We would also like to note that this actually provides us with the Cauchy kernel 
$\Oa_{k,l'}=\frac{1}{2}\frac{k-l'}{k+l'}$ in the linear integral equation. Using Equations \eqref{BKP:CDyn} and \eqref{BKP:Oa}, and differentiating the 
general structure of $\bU$, i.e. \eqref{U}, one immediately obtains the dynamical evolution of $\bU$ with respect to the independent variable $x_j$ given by 
\begin{align}\label{BKP:UDyn}
 \partial_{2j+1}\bU=\bLd^{2j+1}\cdot\bU+\bU\cdot\tbLd^{2j+1}-\frac{1}{2}\bU\cdot(\bO_{2j+1}\cdot\bLd-\tbLd\cdot\bO_{2j+1})\cdot\bU, \quad j=0,1,2,\cdots.
\end{align}
In addition, it can be proven that the infinite matrix $\bU$ 
obeys the antisymmetry property $\tbU=-\bU$. This can be derived from the antisymmetry property of $\bC$ and $\bOa$ (the antisymmetry of $\bOa$ is 
obvious since the kernel is antisymmetric in $k$ and $l'$). We can now refer to \eqref{BKP:UDyn} together with the antisymmetry property of $\bU$ 
as the BKP hierarchy in infinite matrix form because the structure of BKP (i.e. the infinite matrix $\bC$ and the operator $\bOa$) has been contained 
in the two relations. 

In order to get a closed-form scalar equation, we set $u=U_{1,0}=-U_{0,1}$. Like AKP, consider $x_1$-, $x_3$- and $x_{2j+1}$-flows in the equation 
\eqref{BKP:UDyn} and eliminating all the other variables apart from $u$, the nonlinear form of the $x_{2j+1}$-flow of the BKP hierarchy can be obtained. 
Among them the first nontrivial equation is the $x_5$-flow, which is the BKP equation, i.e. 
\begin{align}\label{BKP:NL}
 9u_{x_5x_1}-5u_{x_3x_3}+(-5u_{x_1x_1x_3}-15u_{x_1}u_{x_3}+u_{x_1x_1x_1x_1x_1}+15u_{x_1}u_{x_1x_1x_1}+15u_{x_1}^3)_{x_1}=0.
\end{align}
Other nonlinear forms (e.g. the modified BKP equation, etc.) in the BKP class also exist like AKP, but multi-component forms must be involved. 
We omit them here because we would prefer to be focusing on scalar nonlinear forms. 

The $\tau$-function in the BKP class is defined as $\tau^2=\det(1+\bOa\cdot\bC)$. The reason why we define $\tau^2$ here instead of $\tau$ is because 
the determinant must be in the form of a perfect square due to the antisymmetry property of $\bOa$ and $\bC$ and this treatment will lead to the 
equations in $\tau$ written in a more elegant form. Some direct computation yields the dynamical evolution of $\partial_{2j+1}\ln\tau^2$ and the 
simplest one gives rise to the transform $(\ln\tau^2)_{x_1}=2(\ln\tau)_{x_1}=u$, which transfers the nonlinear BKP hierarchy to its multilinear form. 
The first nontrivial multilinear equation in the hierarchy can be obtained from \eqref{BKP:NL}, which is given by
\begin{align}\label{BKP:BL}
 (D_1^6-5D_1^3D_3-5D_3^2+9D_1D_5)\tau\cdot\tau=0,
\end{align}
where $D_i$ is Hirota's operator. This is the bilinear form of the nonlinear BKP equation \eqref{BKP:NL}. 

Now we consider the linear problem for the BKP hierarchy. In fact, differentiating \eqref{uk} with respect to $x_{2j+1}$, one has the dynamical evolution 
\begin{align}\label{BKP:ukDyn}
 \partial_{2j+1}\bu_k=\bLd^{2j+1}\cdot\bu_k-\frac{1}{2}\bU\cdot(\bO_{2j+1}\cdot\bLd-\tbLd\cdot\bO_{2j+1})\cdot\bu_k, \quad j=0,1,2,\cdots,
\end{align}
where \eqref{BKP:UDyn} and \eqref{BKP:ukDyn} are used. One can refer to this relation as the linear problem for the BKP hierarchy in infinite matrix 
form. Similarly to the linear problem for AKP, we denote $\phi=u_k^{(0)}$ and after eliminating the other components $u_k^{(i)}$ for $i\neq0$ in 
Equation \eqref{BKP:ukDyn}, one can get the linear problems for the whole BKP hierarchy. For instance the spectral problem and the temporal evolution 
for the BKP equation \eqref{BKP:NL} are given by 
\bse\label{BKP:Lax}
\begin{align}
 &\phi_{x_3}=(\partial_1^3+3u_{x_1}\partial_1)\phi, \label{BKP:LaxS} \\
 &\phi_{x_5}=\Big[\partial_1^5+5u_{x_1}\partial_1^3+5u_{x_1x_1}\partial_1^2
 +\Big(\frac{10}{3}u_{x_1x_1x_1}+5u_{x_1}^2+\frac{5}{3}u_{x_3}\Big)\partial_1\Big]\phi. \label{BKP:LaxT}
\end{align}
\ese
The linear problem for \eqref{BKP:BL} can be obtained using the transform $u=2(\ln\tau)_{x_1}$. We note that we made use of the dynamical evolution and 
the antisymmetry property of the infinite matrix of $\bU$ in the derivation. The temporal parts for the other members in the hierarchy can be calculated 
in the same way by considering the corresponding flows in \eqref{BKP:ukDyn}.

\subsection{The CKP hierarchy}\label{S:CKP}
For the KP hierarchy of C-type (associated with the infinite-dimensional Lie algebra $C_{\infty}$), we define the infinite matrix $\bC$ as
\begin{align}\label{CKP:C}
 \bC=\iint_D\rd\zeta(l,l')\rho_l\bc_l\tbc_{l'}\rho_{l'}, \quad \rho_k=\exp\Big(\sum_{j=0}^\infty k^{2j+1}x_{2j+1}\Big), \quad 
 \rd\zeta(l',l)=\rd\zeta(l,l').
\end{align}
From the definition one can see that the only difference between the BKP and CKP classes is that the measure is now symmetric in the CKP case. 
The symmetric measure immediately provides the property $\bC=\tbC$. Since there are only odd flows in the plane wave factors in the $\bC$, 
the dynamical evolutions of the infinite matrix $\bC$ only involve the odd flows and they are given by 
\begin{align}\label{CKP:CDyn}
 \partial_{2j+1}\bC=\bLd^{2j+1}\cdot\bC+\bC\cdot\tbLd^{2j+1},
\end{align}
which is exactly the same as that in the BKP class. We now require that the $\bOa$ in the CKP class obeys the algebraic relation 
\begin{align}\label{CKP:Oa}
 \bOa\cdot\bLd+\tbLd\cdot\bOa=\bO.
\end{align}
This relation for the operator $\bOa$ is the same as \eqref{AKP:Oa}, and this tells us that the Cauchy kernel in this case is $\Oa_{k,l'}=\frac{1}{k+l'}$. 
Making use of the symmetry property of $\bC$ and $\bOa$ and following the definition of $\bU$ \eqref{U}, we have the symmetry property $\tbU=\bU$. 
One can now consider the dynamical evolution of $\bU$ with the help of \eqref{CKP:CDyn} and \eqref{CKP:Oa}. By differentiating $\bU$ with respect 
to the independent variables $x_{2j+1}$, we have 
\begin{align}\label{CKP:UDyn}
 \partial_{2j+1}\bU=\bLd^{2j+1}\cdot\bU+\bU\cdot\tbLd^{2j+1}-\bU\cdot\bO_{2j+1}\cdot\bU, \quad j=0,1,2,\cdots.
\end{align}
The dynamical relation \eqref{CKP:UDyn}, together with the symmetry property of $\bU$, can be thought of as the CKP hierarchy in infinite matrix form. 

We now look for scalar closed-form equations from the infinite matrix structure by choosing particular entries in $\bU$. In fact, one can take 
$u=U_{0,0}$ and select $x_1,x_3$ and $x_{2j+1}$ as the independent variables for the $x_{2j+1}$-flow of the CKP hierarchy in \eqref{CKP:UDyn}. 
As a result the CKP hierarchy can be found and the first nontrivial equation is the CKP equation 
\begin{align}\label{CKP:NL}
 9u_{x_5x_1}-5u_{x_3x_3}+\Big(-5u_{x_1x_1x_3}-15u_{x_1}u_{x_3}+u_{x_1x_1x_1x_1x_1}+15u_{x_1}u_{x_1x_1x_1}+15u_{x_1}^3+\frac{45}{4}u_{x_1x_1}^2\Big)_{x_1}=0.
\end{align}
By selecting other entries in $\bU$, one can also obtain other nonlinear forms from the infinite matrix structure nevertheless the price one has to pay 
is that they may be in multi-component form and thus we omit them here. A unified structure describing the CKP class should be the form in the 
$\tau$-function. In this case, it is defined by $\tau=\det(1+\bOa\cdot\bC)$ like that in AKP and therefore it obeys the dynamical evolution 
$\partial_{2j+1}(\ln\tau)=\tr(\bO_j\cdot\bU)$ in which $\bU$ is symmetric. The first one of them gives us the transform $u=(\ln\tau)_{x_1}$ and therefore 
one obtains the multilinear form of the CKP equation \eqref{CKP:NL} given by 
\begin{align}\label{CKP:BL}
 &4\tau^3\tau_{x_1x_1x_1x_1x_1x_1}+5\tau^2\tau_{x_1x_1x_1}^2-24\tau^2\tau_{x_1}\tau_{x_1x_1x_1x_1x_1}-30\tau\tau_{x_1}\tau_{x_1x_1}\tau_{x_1x_1x_1}
 +45\tau_{x_1}^2\tau_{x_1x_1}^2 \nonumber \\
 &+60\tau\tau_{x_1}\tau_{x_1x_1x_1x_1}-60\tau_{x_1}^3\tau_{x_1x_1x_1}+60\tau^2\tau_{x_1}\tau_{x_1x_1x_3}-60\tau\tau_{x_1}^2\tau_{x_1x_3}
 +60\tau_{x_1}^3\tau_{x_3}-60\tau\tau_{x_1}\tau_{x_1x_1}\tau_{x_3} \nonumber \\
 &+20\tau^2\tau_{x_1x_1x_1}\tau_{x_3}-20\tau^3\tau_{x_1x_1x_1x_3}+20\tau^2\tau_{x_3}^2-20\tau^3\tau_{x_3x_3}+36\tau^3\tau_{x_1x_5}-36\tau^2\tau_{x_1}\tau_{x_5}=0.
\end{align}
This quadrilinear form is analogous to the result for the discrete CKP equation (cf. e.g. \cite{FN16}) which is in the form of Cayley's 
$2\times2\times2$ hyperdeterminant. In other words, Equation \eqref{CKP:BL} can be understood as the continuous analogue of a hyperdeterminant. 
The multilinear transform also brings the higher-order equations in $u$ in the hierarchy to the corresponding multilinear forms in the $\tau$-function. 

Similarly to how we derive \eqref{CKP:UDyn}, one can from \eqref{CKP:UDyn} and \eqref{CKP:Oa} obtain the dynamical evolution of $\bu_k$ as follow:
\begin{align}\label{CKP:ukDyn}
 \partial_{2j+1}\bu_k=\bLd^{2j+1}\bu_k-\bU\cdot\bO_{2j+1}\cdot\bu_k, \quad j=0,1,2,\cdots.
\end{align}
One can refer to these relations together with $\bU=\tbU$ as the linear problems for the CKP hierarchy in infinite matrix form. In fact, if we 
fix $x_3$ and $x_1$ and set $\phi=u_k^{(0)}$, the spatial part of the linear problem is derived by getting rid the other components in $\bu_k$, and 
it is given by 
\bse\label{CKP:Lax}
\begin{align}
 \phi_{x_3}=\Big(\partial_1^3+3u_{x_1}\partial_1+\frac{3}{2}u_{x_1x_1}\Big)\phi. \label{CKP:LaxS}
\end{align}
This is the spatial part for the whole CKP hierarchy. The temporal part can also be calculated from \eqref{CKP:ukDyn} as well. For the CKP equation 
\eqref{CKP:BL}, namely the time variable is fixed at $x_5$, we have the temporal evolution 
\begin{align}
 \phi_{x_5}={}&\Big[\partial_1^5+5u_{x_1}\partial_1^3+\frac{15}{2}u_{x_1x_1}\partial_1^2 \nonumber \\
 &+\Big(\frac{35}{6}u_{x_1x_1x_1}+5u_{x_1}^2+\frac{5}{3}u_{x_3}\Big)\partial_1
 +\Big(\frac{5}{3}u_{x_1x_1x_1x_1}+5u_{x_1}u_{x_1x_1}+\frac{5}{6}u_{x_1x_3}\Big)\Big]\phi. \label{CKP:LaxT}
\end{align}
\ese
The temporal evolutions for the higher-order equations in the hierarchy can be obtained similarly but the formulae of them become more and more complex.

\section{(1+1)-dimensional soliton hierarchies}\label{S:2D}

Dimensional reductions of higher-dimensional hierarchies can always be thought of a powerful tool to obtain (1+1)-dimensional integrable hierarchies. 
In general, following Sato's scheme, such reductions are normally done by imposing certain conditions on pseudo-differential operators and consequently 
the Lax operators for the corresponding lower-dimensional hierarchies arise, cf. e.g. the Kyoto school \cite{MJD00}, Konopelchenko and Strampp 
\cite{KS91,KS92}, Cheng and Li \cite{CL91}, and also Loris and Willox \cite{LW99,Lor99}, etc. Among these dimensional reductions, some of them give 
rise to integrable systems associated with matrix or nonlocal scalar Lax structure which we think will arise from the view point of the DL starting 
from a matrix linear integral equation. We do not discuss these systems in the current paper and only consider the dimensional reductions leading to 
scalar differential spectral problems. 

Dimensional reductions can also be realised within the DL framework. As we can see in the previous section, the integral equation associated with 
(2+1)-dimensional soliton hierarchies is a nonlocal Riemann--Hilbert problem, namely, there must be a double integral in it. When the double integral 
collapses, the integral equation turns out to be a local Riemann--Hilbert problem which is associated with (1+1)-dimensional soliton models. 
This method was recently used in \cite{ZZN12} in order to construct an extended discrete BSQ equation and under such reductions more general solutions 
can be found for the obtained reduced soliton hierarchies. In this section, we generalise this method generically to the AKP, BKP and CKP hierarchies and 
as a result we obtain a huge class of (1+1)-dimensional soliton hierarchies. 

\subsection{Dimensional reductions of the KP-type hierarchies}\label{S:Reduc}

\paragraph{Reductions of the AKP hierarchy.} We take the measure in a particular form as 
\begin{align}\label{AKP:Reduc}
 \rd\zeta(l,l')=\sum_{j=1}^N\rd\ld_j(l)\rd l'\delta(l'+\oa^j l),
\end{align}
where $\omega$ is the $N$th root of unity, namely, $\omega^N=1$. The linear integral equation for the AKP hierarchy with a double integral then 
turns out to be an integral equation with only a single integral as follow: 
\begin{align}\label{A:IntEq}
 \bu_k+\sum_{j=1}^N\int_{\Gamma_j}\rd\ld_j(l)\frac{\rho_k\sigma_{-\oa^jl}}{k-\oa^jl}\bu_l=\rho_k\bc_k, \quad 
 \rho_k=\exp\Big(\sum_{j=1}^\infty k^jx_j\Big), \quad \sigma_{k'}=\exp\Big(-\sum_{j=1}^\infty (-k')^jx_j\Big),
\end{align}
where $\Gamma_j$ is a certain contour (as degeneration of the domain $D$) for the integral and $\rd\ld_j(l)$ is the associated measure. This integral 
equation is the one for the $N$th member in the Gel'fand--Dikii (GD) hierarchy. In fact, such a reduction imposed on the AKP hierarchy also gives 
reduced forms of the infinite matrix $\bC$ and the infinite matrix $\bU$ respectively: 
\begin{align}\label{A:U}
 \bC=\sum_{j=1}^N\int_{\Gamma_j}\rd\ld_j(l)\rho_l\bc_l\tbc_{-\oa^j l}\sigma_{-\oa^j l}, \quad 
 \bU=\sum_{j=1}^N\int_{\Gamma_j}\rd\ld_j(l)\bu_l\tbc_{-\oa^jl}\sigma_{-\oa^jl}.
\end{align}
Nevertheless, one can easily verify the representation of $\bU$ in terms of the infinite matrices \eqref{U} will be invariant, i.e. we still have 
$\bU=(1-\bU\cdot\bOa)\cdot\bU$, and so is that of $\bu_k$ \eqref{uk}, namely, $\bu_k=(1-\bU\cdot\bOa)\cdot\bu_k$. The same thing will also happen 
in the dimensional reductions of the BKP and CKP hierarchies. 

Consider the property of the $N$th root of unity, we can obviously see from the structure of the $\bC$, i.e. Equation \eqref{A:U}, that $\bC$ satisfies 
not only the dynamical relation \eqref{AKP:CDyn} but also the following algebraic relation: 
\begin{align}\label{A:CAlg}
 \partial_j\bC=\bLd^j\cdot\bC-(-\tbLd)^j\cdot\bC=0, \quad j=0 \mod N.
\end{align}
In other words, from the view point of the DL framework, some dynamical relations degenerate to algebraic relations. This observation immediately leads 
to a similar property for the nonlinear variable $\bU$ which is given by
\begin{align}\label{A:UAlg}
 \partial_j\bU=\bLd^j\cdot\bU-\bU\cdot(-\tbLd)^j-\bU\cdot\bO_j\cdot\bU=0, \quad j=0 \mod N.
\end{align}
In fact, this algebraic relation can be proven by differentiating \eqref{U} with respect to $x_j$ and making use of \eqref{AKP:Oa} (which is invariant 
under the reduction), \eqref{AKP:CDyn} and \eqref{A:CAlg}. Equations \eqref{AKP:UDyn} together with \eqref{A:UAlg} can be considered as the 
hierarchy in GD of rank $N$ in infinite matrix form. Furthermore, for the multilinear variable $\tau$ one can easily derive the constraints 
$\partial_{j}\tau=0$ for $j=0 \mod N$ due to \eqref{A:U} and the variable $\bu_k$ for the linear problem now obeys the algebraic relation 
\begin{align}\label{A:ukAlg}
 \partial_j\bu_k=\bLd^j\cdot\bu_k-\bU\cdot\bO_j\cdot\bu_k=k^j\bu_k, \quad j=0 \mod N,
\end{align}
which together with \eqref{AKP:ukDyn} will provide the Lax pairs for the reduced (1+1)-dimensional hierarchies.

\paragraph{Reductions of the BKP hierarchy.} In the BKP class, we introduce the reduction of the measure in a slightly different way, which is given by
\begin{align}\label{BKP:Reduc}
 \rd\zeta(l,l')=\sum_{j=1}^N\rd\ld_j(l)\rd l'\delta(l'+\oa^j l)-\sum_{j=1}^N\rd\ld_j(l')\rd l\delta(l+\oa^j l').
\end{align}
The reason why we take this more complex form compared to \eqref{AKP:Reduc} is because in the BKP hierarchy the measure $\rd\zeta(l,l')$ is antisymmetric 
and only the dimensional reduction \eqref{BKP:Reduc} can preserve the property. Like the reductions of AKP, we have that in the BKP hierarchy 
the reduced integral equation is given by 
\begin{align}\label{B:IntEq}
 \bu_k+\sum_{j=1}^N\int_{\Gamma_j}\rd\ld_j(l)\frac{1}{2}\rho_k\frac{k+\oa^jl}{k-\oa^jl}\rho_{-\oa^jl}\bu_l
 -\sum_{j=1}^N\int_{\Gamma_j}\rd\ld_j(l')\frac{1}{2}\rho_k\frac{k-l'}{k+l'}\rho_{l'}\bu_{-\oa^jl}=\rho_k\bc_k.
\end{align}
And the reduced infinite matrix $\bC$ becomes 
\begin{align}\label{B:C}
 \bC=\sum_{j=1}^N\int_{\Gamma_j}\rd\ld_j(l)\rho_l\bc_l\tbc_{-\oa^j l}\rho_{-\oa^j l}
 -\sum_{j=1}^N\int_{\Gamma_j}\rd\ld_j(l')\rho_{-\oa^j l'}\bc_{-\oa^j l'}\tbc_{l'}\rho_{l'},
\end{align}
where $\rho_k=\exp(\sum_{j=0}^\infty k^{2j+1}x_{2j+1})$ and obviously it still obeys $\tbC=-\bC$. Simultaneously, for the nonlinear 
variable $\bU$ we have 
\begin{align}\label{B:U}
 \bU=\sum_{j=1}^N\int_{\Gamma_j}\rd\ld_j(l)\bu_l\tbc_{-\oa^jl}\rho_{-\oa^jl}-\sum_{j=1}^N\int_{\Gamma_j}\rd\ld_j(l')\bu_{-\oa^jl'}\tbc_{l'}\rho_{l'},
\end{align}
which also obeys the antisymmetry property $\tbU=-\bU$ following from that of $\bOa$ defined in \eqref{BKP:Oa} and that of the reduced $\bC$. 

Some straightforward computation shows that the infinite matrix $\bC$ in the $N$-reduction of the BKP hierarchy obeys 
\begin{align}\label{B:CAlg}
 \bLd^j\cdot\bC-(-\tbLd)^j\cdot\bC=0, \quad j=0 \mod N,
\end{align}
and this together with \eqref{U} and \eqref{BKP:Oa} give rise to the algebraic relation for the nonlinear variable $\bU$ which is given by 
\begin{align}\label{B:UAlg}
 \bLd^j\cdot\bU-\bU\cdot(-\tbLd)^j-\frac{1}{2}\bU\cdot(\bO_j\cdot\bLd-\tbLd\cdot\bO_j)\cdot\bU=0, \quad j=0 \mod N.
\end{align}
This relation obviously implies that $\partial_{2j+1}\bU=0$ when $2j+1=0\mod N$ in \eqref{BKP:UDyn} and consequently we have also $\partial_{2j+1}\tau=0$. 
Similarly one can also from \eqref{uk} prove that 
the reduction \eqref{BKP:Reduc} gives us 
\begin{align}
 k^j\bu_k=\bLd^j\cdot\bu_k-\frac{1}{2}\bU\cdot(\bO_j\cdot\bLd-\tbLd\cdot\bO_j)\cdot\bu_k, \quad j=0 \mod N,
\end{align}
and particularly when $2j+1=0\mod N$ this algebraic relation implies $\partial_{2j+1}\bu_k=k^{2j+1}\bu_k$ if one follows from \eqref{BKP:ukDyn}. 
This relation together with \eqref{BKP:ukDyn} will later give us the Lax pairs for the (1+1)-dimensional hierarchies arising from the reductions 
of the BKP class. 

\paragraph{Reductions of the CKP hierarchy.} While in the CKP class, we take a particular measure in the form of 
\begin{align}\label{CKP:Reduc}
 \rd\zeta(l,l')=\sum_{j=1}^N\rd\ld_j(l)\rd l'\delta(l'+\oa^j l)+\sum_{j=1}^N\rd\ld_j(l')\rd l\delta(l+\oa^j l').
\end{align}
This reduction on the measure has been symmetrised, namely, the reduced measure preserves the symmetry property, and it in turn implies that 
we now have the integral equation for the reduced hierarchies from the CKP class as 
\begin{align}\label{C:IntEq}
 \bu_k+\sum_{j=1}^N\int_{\Gamma_j}\rd\ld_j(l)\frac{\rho_k\rho_{-\oa^jl}}{k-\oa^jl}\bu_l
 +\sum_{j=1}^N\int_{\Gamma_j}\rd\ld_j(l')\frac{\rho_k\rho_{l'}}{k+l'}\bu_{-\oa^jl}=\rho_k\bc_k,
\end{align}
where $\rho_k$ is exactly the same as the one in the reductions of BKP, namely $\rho_k=\exp(\sum_{j=0}^\infty k^{2j+1}x_{2j+1})$ . 
The infinite matrix $\bC$ in this case is now expressed by 
\begin{align}\label{C:C}
 \bC=\sum_{j=1}^N\int_{\Gamma_j}\rd\lambda_j(l)\rho_l\bc_l\tbc_{-\oa^j l}\rho_{-\oa^j l}
 +\sum_{j=1}^N\int_{\Gamma_j}\rd\ld_j(l')\rho_{-\oa^j l'}\bc_{-\oa^j l'}\tbc_{l'}\rho_{l'},
\end{align}
and it obeys the symmetry property $\tbC=\bC$. Furthermore the nonlinear variable $\bU$ now turns out to be 
\begin{align}\label{C:U}
 \bU=\sum_{j=1}^N\int_{\Gamma_j}\rd\ld_j(l)\bu_l\tbc_{-\oa^jl}\rho_{-\oa^jl}+\sum_{j=1}^N\int_{\Gamma_j}\rd\ld_j(l')\bu_{-\oa^jl'}\tbc_{l'}\rho_{l'},
\end{align}
and due to the symmetry property of $\bC$ and $\bOa$ in the reductions, from \eqref{U} one still has $\tbU=\bU$. 

Notice that $\oa^N=1$, one can easily prove that the infinite matrix $\bC$ obeys the algebraic relation 
\begin{align}\label{C:CAlg}
 \bLd^j\cdot\bC-(-\tbLd)^j\cdot\bC=0, \quad j=0 \mod N,
\end{align}
which is exactly the same as the one in the reductions of BKP, i.e. \eqref{B:CAlg}, and consequently one can now follow from \eqref{U} and obtain the 
important algebraic relation for the nonlinear dynamical variable $\bU$ as follow: 
\begin{align}\label{C:UAlg}
 \bLd^j\cdot\bU-\bU\cdot(-\tbLd)^j-\bU\cdot\bO_j\cdot\bU=0, \quad j=0 \mod N.
\end{align}
In the particular case when $2j+1=0\mod N$, this above algebraic relation yields $\partial_{2j+1}\bU=0$. While from \eqref{C:CAlg} we can prove 
$\partial_{2j+1}\bC=0$ and consequently $\partial_{2j+1}\tau=0$ in the same case. 

On the linear level, taking \eqref{C:UAlg} into consideration, one can find from \eqref{uk} that the linear variable $\bu_k$ satisfies the algebraic relation 
\begin{align}\label{C:ukAlg}
 k^j\bu_k=\bLd^j\cdot\bu_k-\bU\cdot\bO_j\cdot\bu_k, \quad j=0 \mod N,
\end{align}
which implies $\partial_{2j+1}\bu_k=k^{2j+1}\bu_k$ when $2j+1=0\mod N$ and this relation is the ingredient for Lax pairs of the reduced lower-dimensional 
equations. 

Dimensional reductions of the KP-type equations are associated with finite-dimensional Kac--Moody Lie algebras as sub-algebras of $A_\infty$, $B_\infty$ 
and $C_{\infty} $\cite{JM83}. Table \ref{Tab:Reduc} shows the first few examples of the dimensional reductions of the KP-type equations. One remark is that 
the 2-reduction of the BKP hierarchy leads to triviality due to the antisymmetry property of the measure $\rd\zeta(l,l')$. In the following subsections, 
we give the DL scheme for the hierarchies of all the equations listed in the table as examples according to the generic scheme for the dimensional reductions 
given in this subsection. 
\begin{table}
\begin{center}
\begin{tabular}{l|l|l}
\hline
Dimensional reductions & Affine Lie algebras & (1+1)-dimensional soliton hierarchies \\
\hline
2-reduction of AKP & $A_1^{(1)}$ & Korteweg--de Vries \\
3-reduction of AKP & $A_2^{(1)}$ & Boussinesq \\
4-reduction of AKP & $A_3^{(1)}$ & generalised Hirota--Satsuma \cite{SH82} \\
\hline
3-reduction of BKP/CKP & $A_2^{(2)}$ & Sawada--Kotera \cite{SK74} and Kaup--Kupershmidt \cite{Kau80} \\
5-reduction of BKP/CKP & $A_4^{(2)}$ & bidirectional SK and bidirectional KK \cite{DP01} \\
\hline
2-reduction of CKP & $C_1^{(1)}\approx A_1^{(1)}$ & Korteweg--de Vries \\
4-reduction of CKP & $C_2^{(1)}$ & Hirota--Satsuma \cite{HS81,SH82,DS85}\\
\hline
4-reduction of BKP & $D_2^{(2)}\approx A_1^{(1)}$ & Korteweg--de Vries \\
6-reduction of BKP & $D_3^{(2)}$ & Ito \cite{Ito80} \\
\hline
\end{tabular}
\caption{Some examples for the dimensional reductions of AKP, BKP and CKP}
\label{Tab:Reduc}
\end{center}
\end{table}

\subsection{The Korteweg--de Vries hierarchy}\label{S:KdV}

The KdV hierarchy is from the 2-reduction of the AKP hierarchy. The general scheme for KdV is given by the dynamical relation and the algebraic relation
\bse
\begin{align}
 &\partial_{2j+1}\bU=\bLd^{2j+1}\cdot\bU+\bU\cdot\tbLd^{2j+1}-\bU\cdot\bO_{2j+1}\cdot\bU, \quad j=0,1,2,\cdots, \label{KdV:UDyn} \\
 &\bLd^2\cdot\bU-\bU\cdot\tbLd^2-\bU\cdot\bO_2\cdot\bU=0. \label{KdV:UAlg}
\end{align}
\ese
These relations constitute the infinite matrix version of the KdV hierarchy. In order to obtain some closed-form equations, namely some 
scalar equations, from the scheme, we take $u=U_{0,0}$, $v=[\ln(1-U_{0,-1})]_{x_1}$ and $z=U_{-1,-1}-x_1$ and these variables give rise to the 
KdV, modified KdV (mKdV) and Schwarzian KdV (SKdV) hierarchies. The first nontrivial equation (when $j=1$ in \eqref{KdV:UDyn}) in the KdV hierarchy 
is the KdV equation: 
\begin{align}\label{KdV:NL}
 u_{x_3}=\frac{1}{4}u_{x_1x_1x_1}+\frac{3}{2}u_{x_1}^2.
\end{align}
From \eqref{KdV:UDyn} and \eqref{KdV:UAlg} one can also find the connection between $u$ and $v$ as well as that between $v$ and $z$, i.e. the 
Miura-type transforms among the nonlinear variables given by 
\begin{align}\label{KdV:MT}
 u_{x_1}=-\frac{1}{2}(v_{x_1}+v^2), \quad v=\frac{1}{2}\frac{z_{x_1x_1}}{z_{x_1}}.
\end{align}
The above transforms help us to derive other nonlinear forms in the KdV class from \eqref{KdV:NL} and we find 
\begin{align}
 &v_{x_3}=\frac{1}{4}v_{x_1x_1x_1}-\frac{3}{2}v^2v_{x_1}, \label{KdV:Mod} \\
 &\frac{z_{x_3}}{z_{x_1}}=\frac{1}{4}\{z,x\}\doteq\frac{1}{4}\frac{z_{x_1x_1x_1}}{z_{x_1}}-\frac{3}{8}\frac{z_{x_1x_1}^2}{z_{x_1}^2}. \label{KdV:Sch}
\end{align}
The equation in $v$ is the mKdV equation and the equation in $z$ is the SKdV equation. Actually, since the 2-reduction of AKP is equivalent to 
$\partial_{2}\bU=0$ as we have pointed out in Subsection \ref{S:Reduc}, the above results can very easily be obtained from the results in the 
AKP class discussed in Subsection \ref{S:AKP}. The higher-order equations in the mKdV and SKdV hierarchies can be calculated via the same Miura-type 
transforms. 

The $\tau$-function of the KdV class is the same as the one for the AKP class with an additional constraint $\partial_{2j}\tau=0$. Therefore the same 
transform $u=(\ln\tau)_{x_1}$ substituted into \eqref{KdV:NL} gives us the bilinear KdV equation 
\begin{align}\label{KdV:BL}
 (D_1^4-4D_1D_3)\tau\cdot\tau=0,
\end{align}
which can also be obtained from \eqref{AKP:BL} directly by reduction. 

The infinite matrix version of the linear problem of the KdV hierarchy can be obtained from the reduction of that for the AKP hierarchy and it is given by 
\begin{align}\label{KdV:ukDyn}
 k^2\bu_k=\bLd^2\cdot\bu_k-\bU\cdot\bO_2\cdot\bu_k, \quad \partial_{2j+1}\bu_k=\bLd^{2j+1}\cdot\bu_k-\bU\cdot\bO_{2j+1}\cdot\bu_k, \quad j=0,1,2,\cdots.
\end{align}
By taking $\phi=(\bu_k)_0$, one can obtain the Lax pair for the KdV hierarchy and the one for the KdV equation \eqref{KdV:NL} is 
\bse\label{KdV:Lax}
\begin{align}
 &L^{\mathrm{KdV}}\phi=k^2\phi, \quad L^{\mathrm{KdV}}=\partial_1^2+2u_{x_1}, \label{KdV:LaxS} \\
 &\phi_{x_3}=\Big(\partial_1^3+3u_{x_1}\partial_1+\frac{3}{2}u_{x_1x_1}\Big)\phi. \label{KdV:LaxT}
\end{align}
\ese
Equation \eqref{KdV:LaxS} is the spectral problem for the whole KdV hierarchy. The Lax pair for the other nonlinear forms \eqref{KdV:Mod} and 
\eqref{KdV:Sch} and the bilinear form \eqref{KdV:BL} can be written down if one replaces $u$ by the other variables according the transforms given 
in this subsection. 

The 2-reduction of the CKP hierarchy is exactly the same as the scheme from that of AKP, therefore the same results can be obtained. 
In fact, the additional constraint $\tbU=\bU$ from CKP (which AKP does not have) only appears in the KdV class ($N=2$) as a special case, 
and it is equivalent to the algebraic relation \eqref{KdV:UAlg}.

The 4-reduction of the BKP hierarchy is slightly different and the infinite matrix structure of $\bU$ is given by 
\begin{align*}
 &\partial_{2j+1}\bU=\bLd^{2j+1}\cdot\bU+\bU\cdot\tbLd^{2j+1}-\frac{1}{2}\bU\cdot(\bO_{2j+1}\cdot\bLd-\tbLd\cdot\bO_{2j+1})\cdot\bU, \quad j=0,1,2,\cdots, \\
 &\bLd^4\cdot\bU-\bU\cdot\tbLd^4-\frac{1}{2}\bU\cdot(\bO_4\cdot\bLd-\tbLd\cdot\bO_4)\cdot\bU=0, \quad \tbU=-\bU.
\end{align*}
Choosing the entry $u=U_{1,0}$ in the infinite matrix one can derive the KdV hierarchy and the first nontrivial equation is 
$u_{x_3}+\frac{1}{2}u_{x_1x_1x_1}+\frac{3}{2}u_{x_1}^2=0$ which is the same as \eqref{KdV:NL} up to some scaling on the independent variables and the 
dependent variable. The bilinear transform $u=2(\ln\tau)_{x_1}$ according to BKP gives us the bilinear form $(D_1^4+2D_1D_3)\tau\cdot\tau=0$ obviously. 
The infinite matrix structure for the linear variable $\bu_k$ in this case obeys 
\begin{align*}
 &k^4\bu_k=\bLd^4\cdot\bu_k-\frac{1}{2}\bU\cdot(\bO_4\cdot\bLd-\tbLd\cdot\bO_4)\cdot\bu_k, \\
 &\partial_{2j+1}\bu_k=\bLd^{2j+1}\cdot\bU+\bU\cdot\tbLd^{2j+1}-\frac{1}{2}\bU\cdot(\bO_{2j+1}\cdot\bLd-\tbLd\cdot\bO_{2j+1})\cdot\bU,
\end{align*}
where $j=0,1,2,\cdots$, and it leads to a 4th-order Lax pair for KdV if we take $\phi=u_k^{(0)}$: 
\begin{align*}
 L^{\mathrm{KdV}}\phi=k^4\phi, \quad \phi_{x_3}=(\partial_1^3+3u_{x_1}\partial_1)\phi,
\end{align*}
where $L^{\mathrm{KdV}}=\partial_1^4+4u_{x_1}\partial_1^2+2u_{x_1x_1}\partial_1$. 

\subsection{The Boussinesq hierarchy}

In this section we consider the 3-reduction of AKP. From the general structure in Subsection \ref{S:Reduc}, when $N=3$ we have 
the dynamical relation for $\bU$ together with the associated algebraic relation as follows: 
\bse
\begin{align}
 &\partial_j\bU=\bLd^j\cdot\bU-\bU\cdot(-\tbLd)^j-\bU\cdot\bO_j\cdot\bU, \quad j\neq 0\mod 3, \quad j\in\mathbb{Z}^+, \label{BSQ:UDyn} \\
 &\bLd^3\cdot\bU+\bU\cdot\tbLd^3-\bU\cdot\bO_3\cdot\bU=0. \label{BSQ:UAlg}
\end{align}
\ese
The relations constitute the infinite matrix representation of the BSQ hierarchy. In fact, these relations are equivalent to the dynamical relations 
for the AKP hierarchy \eqref{AKP:UDyn} subject to a constraint $\partial_j\bU=0$ for $j=0\mod 3$. If one introduces the nonlinear variables 
$u=U_{0,0}$, $v=[\ln(1-U_{0,-1})]_{x_1}$ and $z=U_{-1,-1}-x_1$, from the \eqref{BSQ:UDyn} and \eqref{BSQ:UAlg} one can obtain the BSQ, modified (mBSQ) 
and Schwarzian BSQ (SBSQ) hierarchies respectively. The first nontrivial equation in each hierarchies are 
\begin{align}
 &\Big(\frac{1}{4}u_{x_1x_1x_1}+\frac{3}{2}u_{x_1}^2\Big)_{x_1}+\frac{3}{4}u_{x_2x_2}=0, \label{BSQ:NL} \\
 &\Big(\frac{1}{4}v_{x_1x_1x_1}-\frac{3}{2}v^2v_{x_1}\Big)_{x_1}+\frac{3}{2}v_{x_1x_1}\partial_{x_1}^{-1}v_{x_2}+\frac{3}{2}v_{x_1}v_{x_2}
 +\frac{3}{4}v_{x_2x_2}=0, \label{BSQ:Mod} \\
 &\frac{1}{4}\{z,x_1\}_{x_1} +\frac{3}{4}\frac{z_{x_2}}{z_{x_1}}\Big(\frac{z_{x_2}}{z_{x_1}}\Big)_{x_1}
 +\frac{3}{4}\Big(\frac{z_{x_2}}{z_{x_1}}\Big)_{x_2}=0. \label{BSQ:Sch}
\end{align}
These equations are usually referred to as the BSQ, mBSQ and SBSQ equations respectively. Furthermore, it is obvious to see that the Miura transforms 
for AKP \eqref{AKP:MT} are invariant under the 3-reduction and therefore Equations \eqref{BSQ:NL}, \eqref{BSQ:Mod} and \eqref{BSQ:Sch} are still connected 
by the following Miura-type transforms
\begin{align}\label{BSQ:MT}
 u_{x_1}=-\frac{1}{2}(v_{x_1}+v^2-\partial_{x_1}^{-1}v_{x_2}), \quad  v=\frac{1}{2}\frac{z_{x_1x_1}+z_{x_2}}{z_{x_1}}.
\end{align}

One can now consider the bilinear form of the BSQ equation. In fact, we have already known in Subsection \ref{S:Reduc} that in the BSQ class 
$\tau_{x_3}=0$ and hence from the bilinear KP equation \eqref{AKP:BL} one immediately obtains the bilinear form 
\begin{align}\label{BSQ:BL}
 (D_1^4+3D_2^2)\tau\cdot\tau=0.
\end{align}
Alternatively it can also be obtained by replacing $u$ in \eqref{BSQ:NL} by $(\ln\tau)_{x_1}$ which holds for AKP and its dimensional reductions. 
The multilinear form for the other members in the BSQ hierarchy can be calculated in the same way. 

The linear problem for the BSQ hierarchy in infinite matrix structure is given by 
\begin{align}\label{BSQ:ukDyn}
 k^3\bu_k=\bLd^3\cdot\bu_k-\bU\cdot\bO_3\cdot\bu_k, \quad 
 \partial_{j}\bu_k=\bLd^{j}\cdot\bu_k-\bU\cdot\bO_{j}\cdot\bu_k, \quad j\neq 0 \mod 3, \quad j\in \mathbb{Z}^+.
\end{align}
Eliminating the index-raising operator $\bLd$ and setting $\phi=u_k^{(0)}$, we can derive the Lax pair for the BSQ hierarchy. For instance the Lax 
pair for the $x_2$-flow, i.e. the BSQ equation, is in the form of 
\bse\label{BSQ:Lax}
\begin{align}
 &L^{\mathrm{BSQ}}\phi=k^3\phi, \quad L^{\mathrm{BSQ}}=\partial_1^3+3u_{x_1}\partial_1+\frac{3}{2}(u_{x_1x_1}+u_{x_2}), \label{BSQ:LaxS} \\
 &\phi_{x_2}=(\partial_1^2+2u_{x_1})\phi. \label{BSQ:LaxT}
\end{align}
\ese
The bilinear transform and the Miura-type transforms substituted into the formulae yields the Lax pairs for the modified and Schwarzian hierarchies. 

\subsection{The generalised Hirota--Satsuma hierarchy}

We now consider $N=4$ in the dimensional reductions of the AKP class. In fact, the (1+1)-dimensional integrable hierarchy in this case can longer be 
written in scalar form. In other words, we will obtain a hierarchy of multicomponent systems. Later one can see that the first nontrivial system is 
a 3-component generalisation of the famous HS hierarchy (a coupled KdV hierarchy) and here we refer to it as the generalised HS (gHS) hierarchy. 
According to the general framework, we can write down the infinite matrix formula of $\bU$, namely, the dynamical relation and the algebraic relation 
for $\bU$ given by 
\bse
\begin{align}
 &\partial_j\bU=\bLd^j\cdot\bU-\bU\cdot(-\tbLd)^j-\bU\cdot\bO_j\cdot\bU, \quad j\neq 0\mod 4, \quad j\in\mathbb{Z}^+, \label{gHS:UDyn} \\
 &\bLd^4\cdot\bU-\bU\cdot\tbLd^4-\bU\cdot\bO_4\cdot\bU=0, \label{gHS:UAlg}
\end{align}
\ese
which can be understood as the gHS hierarchy in infinite matrix form. Now we introduce the following nonlinear variables which are some combinations 
of the entries in the infinite matrix $\bU$: 
\begin{align*}
 &u=U_{0,0}, \quad v=[\ln(1-U_{0,-1})]_{x_1}, \quad w=-\frac{U_{1,-1}}{1-U_{0,-1}}, \\
 &r=U_{1,0}-U_{0,1}, \quad s=U_{1,0}+\frac{U_{0,0}U_{1,-1}-U_{2,-1}}{1-U_{0,-1}}, \\
 &p=U_{3,0}+U_{0,3}-U_{2,1}-U_{1,2}+2(U_{0,0}U_{1,1}-U_{1,0}U_{0,1}), \\
 &q=U_{3,0}-U_{0,3}+U_{2,1}-U_{1,2}+2\frac{U_{1,-1}(U_{2,0}-U_{0,2})-U_{2,-1}(U_{1,0}-U_{0,1})}{1-U_{0,-1}}.
\end{align*}
Taking $u$, $p$ and $r$ as the dependent variables of a system, one can from the infinite matrix structure, i.e. \eqref{gHS:UDyn} and \eqref{gHS:UAlg}, 
find the gHS hierarchy. The first nontrivial equation is the $x_3$-flow, namely, the gHS equation 
\bse\label{gHS:NL}
\begin{align}
 &u_{x_3}=\frac{1}{4}u_{x_1x_1x_1}+\frac{3}{2}u_{x_1}^2+\frac{3}{4}(p-r^2), \\
 &p_{x_3}=-\frac{1}{2}p_{x_1x_1x_1}-3u_{x_1}p_{x_1}, \\
 &r_{x_3}=-\frac{1}{2}r_{x_1x_1x_1}-3u_{x_1}r_{x_1}.
\end{align}
\ese
This equation was first given in \cite{SH82} as the 4-reduction on the pseudo-differential operator in the KP hierarchy in order to derive the HS equation. 
In Subsection \ref{S:HS} we will see that this is very natural from the view point of the DL framework. The infinite matrix structure also provides us 
with a Miura transform as follow: 
\begin{align}\label{gHS:MT}
 u_{x_1}=-\frac{1}{2}(v_{x_1}+v^2-s+r), \quad p=-s_{x_1x_1}-s^2-2vs_{x_1}+2sr.
\end{align}
This transform brings us from the gHS hierarchy to its modification. For instance, substituting the transform into the gHS equation gives us the 
generalised modified HS (gmHS) equation 
\bse\label{gHS:Mod}
\begin{align}
 &v_{x_3}=\frac{1}{4}v_{x_1x_1x_1}-\frac{3}{2}v^2v_{x_1}+\frac{3}{2}(vs-vr)_{x_1}+\frac{3}{4}(s+r)_{x_1x_1}, \\
 &s_{x_3}=-\frac{1}{2}s_{x_1x_1x_1}-\frac{3}{2}ss_{x_1}-\frac{3}{2}(v_{x_1}-v^2)s_{x_1}+\frac{3}{2}rs_{x_1}, \\
 &r_{x_3}=-\frac{1}{2}r_{x_1x_1x_1}+\frac{3}{2}rr_{x_1}+\frac{3}{2}(v_{x_1}+v^2)r_{x_1}-\frac{3}{2}sr_{x_1}.
\end{align}
\ese
Furthermore, one can find another Miura transform in the form of 
\begin{align*}
 w_{x_1}=\frac{1}{2}(v_{x_1}-v^2+s-r), \quad q=r_{x_1x_1}-r^2-2vr_{x_1}+2sr,
\end{align*}
and this transform connects the gmHS hierarchy with a hierarchy in terms of the variables $w$, $q$ and $s$. The hierarchy in $w$, $q$ and $s$ is exactly the 
same as the gHS hierarchy. For example the variables obeys the gHS equation \eqref{gHS:NL} if one considers the $x_3$-flow. At the moment we do not need 
this hierarchy as $w$, $q$ and $s$ do not bring us a new nonlinear form for the gHS hierarchy but later in Subsection \ref{S:HS} we will see that the gHS 
hierarchies in $(u,p,r)$ and $(w,q,s)$ will lead to two different nonlinear forms in the C-type reduction. 

Since the gHS is the 4-reduction of AKP, the independent variables apart from $x_{4j}$ for $j\in\mathbb{Z}^+$ still exist in the hierarchy. Notice this 
and consider \eqref{AKP:tau}, one can identify that $u=(\ln\tau)_{x_1}$, $r=(\ln\tau)_{x_2}$ and $p=(\ln\tau)_{x_2x_2}+(\ln\tau)_{x_2}^2$. The multilinear 
transforms help us to reformulate the gHS hierarchy and to get its multilinear form. For instance, we can from \eqref{gHS:NL} derive 
\begin{align}\label{gHS:BL}
 (D_1^4-4D_1D_3+3D_2^2)\tau\cdot\tau=0, \quad (D_1^3D_2+2D_2D_3)\tau\cdot\tau=0, \quad (D_1^3D_2+2D_2D_3)\tau\cdot\tau_{x_2}=0.
\end{align}
The system of bilinear equations is a system with independent variables to $x_1$, $x_2$ and $x_3$. Since there is no $x_2$ in the nonlinear form, here $x_2$ 
can be thought of as an auxiliary variable. If fact, the bilinear form is nothing but the bilinear KP equation in addition to two constraints. 

The linear problem for the gHS class in infinite matrix structure turns out to be 
\begin{align}\label{gHS:ukDyn}
 k^4\bu_k=\bLd^4\cdot\bu_k-\bU\cdot\bO_4\cdot\bu_k, \quad 
 \partial_{j}\bu_k=\bLd^{j}\cdot\bu_k-\bU\cdot\bO_{j}\cdot\bu_k, \quad j\neq 0 \mod 4, \quad j\in \mathbb{Z}^+.
\end{align}
By considering the $x_j$-flows, it gives us the Lax pair for the gHS hierarchy if one sets $\phi=u_k^{(0)}$. Let us take the $x_3$-flow as an example. 
The first equation in \eqref{gHS:ukDyn} gives us the spectral problem and the second one when $j=3$ gives the corresponding temporal part for \eqref{gHS:NL}: 
\bse\label{gHS:Lax}
\begin{align}
 &L^{\mathrm{gHS}}\phi=k^4\phi, \label{gHS:LaxS} \\
 &\phi_{x_3}=\Big[\partial_1^3+3u_{x_1}\partial_1+\frac{3}{2}(u_{x_1x_1}+r_{x_1})\Big]\phi, \label{gHS:LaxT}
\end{align}
\ese
where the differential operator is given by 
\begin{align*}
 L^{\mathrm{gHS}}=\partial_1^4+4u_{x_1}\partial_1^2+(4u_{x_1x_1}+2r_{x_1})\partial_1+2u_{x_1x_1x_1}+4u_{x_1}^2+p+r_{x_1x_1}-r^2.
\end{align*}
The Lax pairs for the hierarchies in other forms such as the modified hierarchy and the bilinear hierarchy can be calculated using the 
transforms given above.

\subsection{The Hirota--Satsuma hierarchy}\label{S:HS}
Now we consider the HS hierarchy as the 4-reduction of the CKP class. The infinite matrix structure in this class is 
\bse\label{HS:UDyn}
\begin{align}
 &\partial_{2j+1}\bU=\bLd^{2j+1}\cdot\bU+\bU\cdot\tbLd^{2j+1}-\bU\cdot\bO_{2j+1}\cdot\bU, \quad j=0,1,2\cdots, \\
 &\bLd^4\cdot\bU-\bU\cdot\tbLd^4-\bU\cdot\bO_4\cdot\bU=0, \quad \tbU=\bU. \label{HS:UAlg}
\end{align}
\ese
If one compares it with the structure in the gHS class, it is obvious to see that the only difference between HS and gHS is that in the HS class one 
only considers odd flows and in addition the infinite matrix $\bU$ satisfies the symmetry property. Therefore one can define the nonlinear variables 
exactly the same as those in gHS, nevertheless some of the variables in gHS now become zero due to the symmetry property. Concretely, we have $r=q=0$. 
The nonlinear variables in the class are now given by 
\begin{align*}
 &u=U_{0,0}, \quad v=[\ln(1-U_{0,-1})]_{x_1}, \quad w=-\frac{U_{1,-1}}{1-U_{0,-1}}, \quad z=U_{-1,-1}-x_1, \\
 &p=2U_{3,0}-2U_{2,1}+2(U_{0,0}U_{1,1}-U_{1,0}^2), \quad s=U_{1,0}+\frac{U_{0,0}U_{1,-1}-U_{2,-1}}{1-U_{0,-1}},
\end{align*}
The hierarchy based on $u$ and $p$ can be obtained from the gHS hierarchy in $(u,p,r)$ by setting $r=0$ directly. The first nontrivial nonlinear 
equation in the hierarchy is 
\bse\label{HS:NL}
\begin{align}
 &u_{x_3}=\frac{1}{4}u_{x_1x_1x_1}+\frac{3}{2}u_{x_1}^2+\frac{3}{4}p, \\
 &p_{x_3}=-\frac{1}{2}p_{x_1x_1x_1}-3u_{x_1}p_{x_1},
\end{align}
\ese
which was first given by Satsuma and Hirota as an equivalent form of the HS equation (cf. \cite{SH82}). While setting $q=0$ in the gHS hierarchy 
expressed by $(w,q,s)$, one can easily find the original form of the HS hierarchy and the HS equation is 
\bse\label{HS:HS}
\begin{align}
 &w_{x_3}=\frac{1}{4}w_{x_1x_1x_1}+\frac{3}{2}w_{x_1}^2-\frac{3}{4}s^2, \\
 &s_{x_3}=-\frac{1}{2}s_{x_1x_1x_1}-3w_{x_1}s_{x_1}.
\end{align}
\ese
So from the view point of the DL, it is very clear that the HS equation is the 4-reduction of the CKP equation and the treatment that taking one dependent 
variable to be zero in a 3-component HS equation (the gHS equation) given in \cite{SH82} naturally follows from the symmetry property of the infinite 
matrix $\bU$. In other words, this can be understood as the reduction from A-type algebra to C-type algebra. The modified hierarchy can be calculated 
in a similar way (taking $r=0$) in the gmHS hierarchy. The first nontrivial equation is the modified HS (mHS) equation 
\bse\label{HS:Mod}
\begin{align}
 &v_{x_3}=\frac{1}{4}v_{x_1x_1x_1}-\frac{3}{2}v^2v_{x_1}+\frac{3}{2}(vs)_{x_1}+\frac{3}{4}s_{x_1x_1}, \\
 &s_{x_3}=-\frac{1}{2}s_{x_1x_1x_1}-\frac{3}{2}ss_{x_1}-\frac{3}{2}(v_{x_1}-v^2)s_{x_1},
\end{align}
\ese
which was first derived in \cite{JM83} with some misprints in the first equation and the correct form was later given in \cite{WGHZ99} following from 
the Miura transform for \eqref{HS:NL}. In fact, the Miura-type transforms in our framework can be obtained very easily from \eqref{gHS:MT} by imposing 
the symmetry constraint of $\bU$: 
\begin{align}\label{HS:MT}
 u_{x_1}=-\frac{1}{2}(v_{x_1}+v^2-s), \quad p=-s_{x_1x_1}-s^2-2vs_{x_1}, \quad 
 w_{x_1}=\frac{1}{2}(v_{x_1}-v^2+s), \quad v=\frac{1}{2}\frac{z_{x_1x_1}}{z_{x_1}}.
\end{align}
The first two transforms is the Miura transform between \eqref{HS:NL} and \eqref{HS:Mod} which coincides with the result in \cite{WGHZ99}. 
The transform between $w$ and $(v,s)$ reveals the clear link between the mHS equation and the HS equation, which was probably not given before, 
to the best of the authors' knowledge. We note that there is also a ``semi-modification'' of the HS equation by considering the factorisation 
of a 4th-order Lax operator (cf. \cite{BEF95}). The transform between $v$ and $z$ gives rise to the Schwarzian form of the HS class and it in 
multicomponent form is written as 
\bse\label{HS:Sch}
\begin{align}
 &\frac{z_{x_3}}{z_{x_1}}=\frac{1}{4}\{z,x_1\}+\frac{3}{2}s, \\
 &s_{x_3}=-\frac{1}{2}s_{x_1x_1x_1}-\frac{3}{2}ss_{x_1}-\frac{3}{4}s_{x_1}\{z,x_1\}.
\end{align}
\ese
We refer to this equation as the Schwarzian HS (SHS) equation. Eliminating $s$ in the system, we obtain a scalar equation where only $z$ in involved: 
\begin{align}\label{HS:SchScalar}
 2\Big(\frac{z_{x_3}}{z_{x_1}}-\frac{1}{4}\{z,x_1\}\Big)_{x_3}+\Big(\frac{z_{x_3}}{z_{x_1}}-\frac{1}{4}\{z,x_1\}\Big)_{x_1x_1x_1} 
 +\Big(2\frac{z_{x_3}}{z_{x_1}}+\{z,x_1\}\Big)\Big(\frac{z_{x_3}}{z_{x_1}}-\frac{1}{4}\{z,x_1\}\Big)_{x_1}=0.
\end{align}
This scalar equation was first proposed in \cite{Wei85} as an example which is M\"obius invariant. Equation \eqref{HS:NL} also gives us a scalar 
form in $u$ if we eliminate $p$. The obtained scalar equation is a higher-order equation with respect to $x_3$ given by 
\begin{align}\label{HS:Scalar}
 8u_{x_3x_3}+2u_{x_1x_1x_1x_3}=(u_{x_1x_1x_1x_1x_1}+18u_{x_1}u_{x_1x_1x_1}+9u_{x_1x_1}^2+24u_{x_1}^3)_{x_1},
\end{align}
which is referred to as the bidirectional Hirota--Satsuma (bHS) equation and the explicit form was first written down in \cite{VM03}.

The equation \eqref{HS:Scalar} makes it possible for us to find the multilinear form of the HS class. In fact, in the CKP class we still have the 
transform $u=(\ln\tau)_{x_1}$ and it gives us a quadrilinear equation as follow: 
\begin{align}\label{HS:BL}
 &\tau^3\tau_{x_1x_1x_1x_1x_1x_1}-\tau^2\tau_{x_1x_1x_1}^2+12\tau\tau_{x_1}^2\tau_{x_1x_1x_1x_1}-12\tau_{x_1}^3\tau_{x_1x_1x_1} \nonumber \\
 &+3\tau^2\tau_{x_1x_1}\tau_{x_1x_1x_1x_1}+9\tau_{x_1}^2\tau_{x_1x_1}^2
 -6\tau\tau_{x_1}\tau_{x_1x_1}\tau_{x_1x_1x_1}-6\tau^2\tau_{x_1}\tau_{x_1x_1x_1x_1x_1} \nonumber \\
 &+6\tau^2\tau_{x_1x_1}\tau_{x_1x_3}+6\tau^2\tau_{x_1}\tau_{x_1x_1x_3}
 -12\tau\tau_{x_1}\tau_{x_1x_1}\tau_{x_3}+12\tau_{x_1}^3\tau_{x_3}-12\tau\tau_{x_1}^2\tau_{x_1x_3} \nonumber \\
 &+2\tau^2\tau_{x_1x_1x_1}\tau_{x_3}-2\tau^3\tau_{x_1x_1x_1x_3}+8\tau^2\tau_{x_3}^2-8\tau^3\tau_{x_3x_3}=0,
\end{align}

The formal linear problem for the HS hierarchy in infinite matrix satisfies the following dynamical and algebraic relations: 
\begin{align}\label{HS:ukDyn}
 k^4\bu_k=\bLd^4\cdot\bu_k-\bU\cdot\bO_4\cdot\bu_k, \quad 
 \partial_{2j+1}\bu_k=\bLd^{2j+1}\cdot\bu_k-\bU\cdot\bO_{2j+1}\cdot\bu_k, \quad j=0,1,2,\cdots.
\end{align}
The relations look the same as \eqref{gHS:ukDyn} but in fact $\bU$ now obeys some additional constraints as we have pointed out in \eqref{HS:UDyn}. 
The wave function $\phi=u_k^{(0)}$ as a component in the infinite vector $\bu_k$ gives the Lax pairs for the whole hierarchy. The explicit form for 
the one of Equation \eqref{HS:NL} as an example is given as follow: 
\bse\label{HS:Lax}
\begin{align}
 &L^{\mathrm{HS}}\phi=k^4\phi, \quad 
 L^{\mathrm{HS}}=\partial_1^4+4u_{x_1}\partial_1^2+4u_{x_1x_1}\partial_1+\frac{5}{3}u_{x_1x_1x_1}+2u_{x_1}^2 +\frac{4}{3}u_{x_3}, \label{HS:LaxS} \\
 &\phi_{x_3}=\Big(\partial_1^3+3u_{x_1}\partial_1+\frac{3}{2}u_{x_1x_1}\Big)\phi. \label{HS:LaxT}
\end{align}
\ese
The Lax pairs for the other nonlinear forms and the multilinear form can be calculated simply by the listed differential transforms. 
We note that the Lax pair for \eqref{HS:HS} was first given in \cite{DF82}. 
\subsection{The Sawada--Kotera and Kaup--Kupershmidt hierarchies}

The $(2j+1)$-reduction of the BKP and CKP classes result in the same infinite-dimensional Lie algebra and therefore the obtained (1+1)-dimensional 
integrable hierarchies are the same. In this subsection we consider the 3-reductions of BKP and CKP hierarchies. We start with the 3-reduction of BKP. 
The dynamical relation \eqref{BKP:UDyn} together with \eqref{B:UAlg} for $N=3$ constitute the infinite matrix structure of the SK hierarchy, i.e. 
\bse
\begin{align}
 &\partial_{2j+1}\bU=\bLd^{2j+1}\cdot\bU+\bU\cdot\tbLd^{2j+1}-\frac{1}{2}\bU\cdot(\bO_{2j+1}\cdot\bLd-\tbLd\cdot\bO_{2j+1})\cdot\bU, \label{SK:UDyn} \\
 &\bLd^3\cdot\bU+\bU\cdot\tbLd^3-\frac{1}{2}\bU\cdot(\bO_3\cdot\bLd-\tbLd\cdot\bO_3)\cdot\bU=0, \quad \tbU=-\bU \label{SK:UAlg}
\end{align}
\ese
for $2j+1\neq 0\mod 3$ where $j=0,1,2,\cdots$. These relations constitute the infinite matrix version of the SK hierarchy and they are equivalent to the 
infinite matrix structure of the BKP hierarchy in addition to $\partial_3\bU=0$. One can take $u=U_{1,0}=-U_{0,1}$ and from the BKP hierarchy the SK 
hierarchy can immediately be derived. The first nontrivial equation in the hierarchy is a 5th-order equation 
\begin{align}\label{SK:NL}
 9u_{x_5}+u_{x_1x_1x_1x_1x_1}+15u_{x_1}u_{x_1x_1x_1}+15u_{x_1}^3=0,
\end{align}
which is the SK equation. The bilinear transform in the BKP class shows that $u=2(\ln\tau)_{x_1}$ and consequently one can have the bilinear form of the 
SK hierarchy. Consider the $x_5$-flow, i.e. Equation \eqref{SK:NL}, we obtain a bilinear equation 
\begin{align}\label{SK:BL}
 (D_1^6+9D_1D_5)\tau\cdot\tau=0,
\end{align}
which can also be derived directly from the bilinear BKP equation because it has been shown in Subsection \ref{S:Reduc} that the 3-reduction also implies 
that $\tau_{x_3}=0$ in the BKP hierarchy. The linear problem following from BKP under the 3-reduction constraint turns out to be 
\bse\label{SK:ukDyn}
\begin{align}
 &k^3\bu_k=\bLd^3\cdot\bu_k-\frac{1}{2}\bU\cdot(\bO_3\cdot\bLd-\tbLd\cdot\bO_3)\cdot\bu_k, \\
 &\partial_{2j+1}\bu_k=\bLd^{2j+1}\cdot\bu_k-\frac{1}{2}\bU\cdot(\bO_{2j+1}\cdot\bLd-\tbLd\cdot\bO_{2j+1})\cdot\bu_k
\end{align}
\ese
in infinite matrix structure, where $2j+1\neq 0 \mod 3$, $j=0,1,2,\cdots$. The wave function $\phi=u_k^{(0)}$ gives rise to the Lax pairs for the 
SK hierarchy and the first nontrivial one is 
\bse\label{SK:Lax}
\begin{align}
 &L^{\mathrm{SK}}\phi=k^3\phi, \quad L^{\mathrm{SK}}=\partial_1^3+3u_{x_1}\partial_1, \label{SK:LaxS} \\
 &\phi_{x_5}=\Big[\partial_1^5+5u_{x_1}\partial_1^3+5u_{x_1x_1}\partial_1^2 +\Big(\frac{10}{3}u_{x_1x_1x_1}+5u_{x_1}^2\Big)\partial_1\Big]\phi. \label{SK:LaxT}
\end{align}
\ese
We omit looking for other nonlinear forms here as the 3-reduction of CKP gives the same (1+1)-dimensional soliton hierarchies and it is more convenient to 
find the other forms there. 

Now we consider the 3-reduction of the CKP hierarchy. The infinite matrix structure (namely the dynamical and algebraic relations for $\bU$) is as follow: 
\bse
\begin{align}
 &\partial_{2j+1}\bU=\bLd^{2j+1}\cdot\bU+\bU\cdot\tbLd^{2j+1}-\bU\cdot\bO_{2j+1}\cdot\bU, \label{KK:UDyn} \\
 &\bLd^3\cdot\bU+\bU\cdot\tbLd^3-\bU\cdot\bO_3\cdot\bU=0, \quad \tbU=\bU, \label{KK:UAlg}
\end{align}
\ese
for $2j+1\neq 0 \mod 3$ where $j=0,1,2,\cdots$, which is the infinite matrix version of the hierarchy in the KK and SK class. 
One can now combine the entries in the infinite matrix $\bU$ and introduce the following nonlinear variables: 
\begin{align*}
 u=U_{0,0}, \quad v=[\ln(1-U_{0,-1})]_{x_1}, \quad w=-\frac{U_{1,-1}}{1-U_{0,-1}}, \quad z=U_{-1,-1}-x_1.
\end{align*}
From Equations \eqref{KK:UDyn} and \eqref{KK:UAlg} one can first of all find a hierarchy based on the variable $u$, which is the KK hierarchy. 
As an example in the hierarchy we consider the $x_5$-flow which is the KK equation 
\begin{align}\label{KK:NL}
 9u_{x_5}+u_{x_1x_1x_1x_1x_1}+15u_{x_1}u_{x_1x_1x_1}+15u_{x_1}^3+\frac{45}{4}u_{x_1x_1}^2=0.
\end{align}
Next, one can find that the infinite matrix dynamical and algebraic relations give rise to the following Miura-type transforms: 
\begin{align}\label{KK:MT}
 u_{x_1}=-\frac{2}{3}v_{x_1}-\frac{1}{3}v^2, \quad w_{x_1}=\frac{1}{3}v_{x_1}-\frac{1}{3}v^2, \quad v=\frac{1}{2}\frac{z_{x_1x_1}}{z_{x_1}}.
\end{align}
The first two transforms are the remarkable nonlinear transforms which were proposed by Fordy and Gibbons, cf. \cite{FG80a,FG80b}. 
Making use of them, one can derive the Fordy--Gibbons (FG) and SK hierarchies from KK respectively. The first nontrivial members 
in the respective hierarchy are given by 
\begin{align}
 &9v_{x_5}+v_{x_1x_1x_1x_1x_1}-20vv_{x_1}v_{x_1x_1}-20v^2v_{x_1x_1x_1}-5v_{x_1x_1}^2-5v_{x_1}^3+5v^4v_{x_1}=0, \label{KK:FG} \\
 &9w_{x_5}+w_{x_1x_1x_1x_1x_1}+15w_{x_1}w_{x_1x_1x_1}+15w_{x_1}^3=0. \label{KK:SK}
\end{align}
We note that the equation in $v$ is normally referred to as the FG equation and the equation in $w$ is exactly the same as the SK equation \eqref{SK:NL}.
The FG equation can be thought of as the modification for both of the SK and KK equations. The Miura-type transform between $v$ and $z$ in \eqref{KK:MT} 
provides us with 
\begin{align}\label{KK:Sch}
 9\frac{z_{x_5}}{z_{x_1}}+\{z,x_1\}_{x_1x_1}+\frac{1}{4}\{z,x_1\}^2=0.
\end{align}
This equation is obviously M\"obius invariant. Since there are SK and KK equations as the unmodified equations in the class we refer to this 
equation as the Schwarzian FG (SFG) equation and it was proposed originally in \cite{Wei84}. 

The $\tau$-function defined in the CKP class obeys that $u=(\ln\tau)_{x_1}$ and this transform substituted into the nonlinear form \eqref{KK:NL} 
gives the multilinear form for the SK and KK class, i.e. 
\begin{align}\label{KK:BL}
 &4\tau^3\tau_{x_1x_1x_1x_1x_1x_1}+5\tau^2\tau_{x_1x_1x_1}^2-24\tau^2\tau_{x_1}\tau_{x_1x_1x_1x_1x_1}-30\tau\tau_{x_1}\tau_{x_1x_1}\tau_{x_1x_1x_1}
 +45\tau_{x_1}^2\tau_{x_1x_1}^2 \nonumber \\
 &+60\tau\tau_{x_1}\tau_{x_1x_1x_1x_1}-60\tau_{x_1}^3\tau_{x_1x_1x_1}+36\tau^3\tau_{x_1x_5}-36\tau^2\tau_{x_1}\tau_{x_5}=0. 
\end{align}
The multilinear forms for the higher-order equations in the hierarchy can be derived under the same transform. 

Finally, we consider the linear problem in this class. According to the reduction of the CKP hierarchy. We have the infinite matrix structure 
\bse\label{KK:ukDyn}
\begin{align}
 &k^3\bu_k=\bLd^3\cdot\bu_k-\bU\cdot\bO_3\cdot\bu_k, \\
 &\partial_{2j+1}\bu_k=\bLd^{2j+1}\cdot\bU+\bU\cdot\tbLd^{2j+1}-\bU\cdot\bO_{2j+1}\cdot\bU,
\end{align}
\ese
where $2j+1\neq 0 \mod 3$ for $j=0,1,2,\cdots$. 
By taking $\phi=u_k^{(0)}$, we can eliminate the other components in \eqref{KK:ukDyn} and find a closed-form relation in $\phi$, which is the Lax 
pair for the $x_{2j+1}$-flow in the hierarchy. For instance, when $j=2$, we get the Lax pair for the KK equation \eqref{KK:NL}: 
\bse\label{KK:Lax}
\begin{align}
 &L^{\mathrm{KK}}\phi=k^3\phi, \quad L^{\mathrm{KK}}=\partial_1^3+3u_{x_1}\partial_1+\frac{3}{2}u_{x_1x_1}, \label{KK:LaxS} \\
 &\phi_{x_5}=\Big[\partial_1^5+5u_{x_1}\partial_1^3+\frac{15}{2}u_{x_1x_1}\partial_1^2
 +\Big(\frac{35}{6}u_{x_1x_1x_1}+5u_{x_1}^2\Big)\partial_1
 +\Big(\frac{5}{3}u_{x_1x_1x_1x_1}+5u_{x_1}u_{x_1x_1}\Big)\Big]\phi. \label{KK:LaxT}
\end{align}
\ese
The Miura-type transforms therefore lead us to the Lax pairs for the other nonlinear forms \eqref{KK:FG}, \eqref{KK:SK} and \eqref{KK:Sch} 
and the multilinear transform gives us the Lax pair for the multilinear form \eqref{KK:BL}.

\subsection{Other higher-rank hierarchies}

Other higher-rank (1+1)-dimensional soliton hierarchies can be obtained in a similar way and in those cases more and more nonlinear forms can be found 
in each class. In this subsection we only give some of them as examples. 

We first consider the 5-reduction of the BKP hierarchy. The infinite matrix structure is given by 
\bse
\begin{align}
 &\partial_{2j+1}\bU=\bLd^{2j+1}\cdot\bU+\bU\cdot\tbLd^{2j+1}-\frac{1}{2}\bU\cdot(\bO_{2j+1}\cdot\bLd-\tbLd\cdot\bO_{2j+1})\cdot\bU, \label{bSK:UDyn} \\
 &\bLd^5\cdot\bU+\bU\cdot\tbLd^5-\frac{1}{2}\bU\cdot(\bO_5\cdot\bLd-\tbLd\cdot\bO_5)\cdot\bU=0, \quad \tbU=-\bU \label{bSK:UAlg}
\end{align}
\ese
for $2j+1\neq 0\mod 5$ where $j=0,1,2,\cdots$. Equations \eqref{bSK:UDyn} together with \eqref{bSK:UAlg} constitute the bSK hierarchy in infinite 
matrix. The infinite matrix structure is equivalent to \eqref{BKP:UDyn} in addition to $\partial_{2j+1}\bU=0$ for $2j+1=0\mod 5$.
Like the BKP class, we take the nonlinear variable $u=U_{1,0}$, one can find the nonlinear form of the bSK hierarchy. 
For example, when $j=1$ in \eqref{bSK:UDyn} we find the $x_3$-flow, which is the bSK equation (cf. \cite{DP01}) 
\begin{align}\label{bSK:NL}
 -5u_{x_3x_3}+(-5u_{x_1x_1x_3}-15u_{x_1}u_{x_3}+u_{x_1x_1x_1x_1x_1}+15u_{x_1}u_{x_1x_1x_1}+15u_{x_1}^3)_{x_1}=0.
\end{align}
The bilinear transforms that follows from the BKP hierarchy is $u=2(\ln\tau)_{x_1}$ and it gives us the bilinear form of the bSK equation: 
\begin{align}\label{bSK:BL}
 (D_1^6-5D_1^3D_3-5D_3^2)\tau\cdot\tau=0.
\end{align}
Equation \eqref{bSK:BL} was actually given earlier in \cite{Ram81} in Painlev\'e test for bilinear formalism and therefore this bilinear equation is 
also referred to as the Ramani equation. The higher-order equations in the hierarchy can be obtained by using the same transform. The 5-reduction of 
the linear problem of the BKP hierarchy gives rise the following relations for 
the infinite vector $\bu_k$: 
\bse\label{bSK:ukDyn}
\begin{align}
 &k^5\bu_k=\bLd^5\cdot\bu_k-\frac{1}{2}\bU\cdot(\bO_5\cdot\bLd-\tbLd\cdot\bO_5)\cdot\bu_k, \\
 &\partial_{2j+1}\bu_k=\bLd^{2j+1}\cdot\bU+\bU\cdot\tbLd^{2j+1}-\frac{1}{2}\bU\cdot(\bO_{2j+1}\cdot\bLd-\tbLd\cdot\bO_{2j+1})\cdot\bU,
\end{align}
\ese
where $2j+1\neq 0\mod 5$ for $j=0,1,2,\cdots$. The dynamical relations for $\bu_k$ yield the Lax pairs for the whole bSK hierarchy if one fixes 
the variable $\phi=u_k^{(0)}$. The simplest nontrivial one is the linear problem of the bSK equation \eqref{bSK:NL} whose explicit form is 
\bse\label{bSK:Lax}
\begin{align}
 &L^{\mathrm{bSK}}\phi=k^5\phi, \label{bSK:LaxS} \\
 &\phi_{x_3}=(\partial_1^3+3u_{x_1}\partial_1)\phi, \label{bSK:LaxT}
\end{align}
\ese
where the Lax operator is given by 
\begin{align*}
L^{\mathrm{bSK}}=\partial_1^5+5u_{x_1}\partial_1^3+5u_{x_1x_1}\partial_1^2+(\frac{10}{3}u_{x_1x_1x_1}+5u_{x_1}^2+\frac{5}{3}u_{x_3})\partial_1.
\end{align*}

Likewise, the 5-reduction of the CKP hierarchy has the infinite matrix structure in $\bU$ as follow: 
\bse
\begin{align}
 &\partial_{2j+1}\bU=\bLd^{2j+1}\cdot\bU+\bU\cdot\tbLd^{2j+1}-\bU\cdot\bO_{2j+1}\cdot\bU, \label{bKK:UDyn} \\
 &\bLd^5\cdot\bU+\bU\cdot\tbLd^5-\bU\cdot\bO_5\cdot\bU=0, \quad \tbU=\bU \label{bKK:UAlg}
\end{align}
\ese
for $2j+1\neq 0\mod 5$ where $j=0,1,2,\cdots$. From the structure if one chooses $u=U_{0,0}$, a closed-form hierarchy can be derived. The $x_3$-flow in 
the hierarchy gives us the bKK equation 
\begin{align}\label{bKK:NL}
 -5u_{x_3x_3}+\Big(-5u_{x_1x_1x_3}-15u_{x_1}u_{x_3}+u_{x_1x_1x_1x_1x_1}+15u_{x_1}u_{x_1x_1x_1}+15u_{x_1}^3+\frac{45}{4}u_{x_1x_1}^2\Big)_{x_1}=0,
\end{align}
which was first introduced in \cite{DP01}. In fact, Equations \eqref{HS:Scalar}, \eqref{bSK:NL}, \eqref{bKK:NL} together with KdV6 are the only four 
integrable cases in a general 6th-order equation \cite{KKKSST07}. Nevertheless, KdV6 is very different from the previous three and it does not appear in 
our framework. 

The multilinear transform in the CKP hierarchy is given by $u=(\ln\tau)_{x_1}$ and therefore a quadrilinear form of Equation \eqref{bKK:NL} can be obtained: 
\begin{align}\label{bKK:BL}
 &4\tau^3\tau_{x_1x_1x_1x_1x_1x_1}+5\tau^2\tau_{x_1x_1x_1}^2-24\tau^2\tau_{x_1}\tau_{x_1x_1x_1x_1x_1}-30\tau\tau_{x_1}\tau_{x_1x_1}\tau_{x_1x_1x_1}
 +45\tau_{x_1}^2\tau_{x_1x_1}^2 \nonumber \\
 &+60\tau\tau_{x_1}\tau_{x_1x_1x_1x_1}-60\tau_{x_1}^3\tau_{x_1x_1x_1}+60\tau^2\tau_{x_1}\tau_{x_1x_1x_3}-60\tau\tau_{x_1}^2\tau_{x_1x_3}
 +60\tau_{x_1}^3\tau_{x_3}-60\tau\tau_{x_1}\tau_{x_1x_1}\tau_{x_3} \nonumber \\
 &+20\tau^2\tau_{x_1x_1x_1}\tau_{x_3}-20\tau^3\tau_{x_1x_1x_1x_3}+20\tau^2\tau_{x_3}^2-20\tau^3\tau_{x_3x_3}=0. 
\end{align}
The dynamical and algebraic relations for the infinite vector $\bu_k$ in this case are given by 
\begin{align}\label{bKK:ukDyn}
 k^5\bu_k=\bLd^5\cdot\bu_k-\bU\cdot\bO_5\cdot\bu_k, \quad 
 \partial_{2j+1}\bu_k=\bLd^{2j+1}\cdot\bU+\bU\cdot\tbLd^{2j+1}-\bU\cdot\bO_{2j+1}\cdot\bU
\end{align}
for $2j+1\neq 0 \mod 5$ where $j=0,1,2,\cdots$. Taking $\phi=u_k^{(0)}$, one can calculate the Lax pair for the bKK hierarchy. For the $x_3$-flow, 
namely, the bKK equation \eqref{bKK:NL}, one has the Lax pair
\bse\label{bKK:Lax}
\begin{align}
 &L^{\mathrm{bKK}}\phi=k^5\phi, \label{bKK:LaxS} \\
 &\phi_{x_3}=\Big(\partial_1^3+3u_{x_1}\partial_1+\frac{3}{2}u_{x_1x_1}\Big)\phi, \label{bKK:LaxT}
\end{align}
\ese
where the differential operator is given as 
\begin{align*}
 L^{\mathrm{bKK}}=\partial_1^5+5u_{x_1}\partial_1^3+\frac{15}{2}u_{x_1x_1}\partial_1^2
 +\Big(\frac{35}{6}u_{x_1x_1x_1}+5u_{x_1}^2+\frac{5}{3}u_{x_3}\Big)\partial_1+5u_{x_1}u_{x_1x_1}+\frac{5}{3}u_{x_1x_1x_1x_1}+\frac{5}{6}u_{x_1x_3}.
\end{align*}

Finally we consider the 6-reduction of the BKP hierarchy. We list the dynamical and algebraic relations as follows: 
\begin{align}
 &\partial_{2j+1}\bU=\bLd^{2j+1}\cdot\bU+\bU\cdot\tbLd^{2j+1}-\frac{1}{2}\bU\cdot(\bO_{2j+1}\cdot\bLd-\tbLd\cdot\bO_{2j+1})\cdot\bU, \quad 
 j=0,1,2,\cdots, \label{Ito:UDyn} \\
 &\bLd^6\cdot\bU-\bU\cdot\tbLd^6-\frac{1}{2}\bU\cdot(\bO_6\cdot\bLd-\tbLd\cdot\bO_6)\cdot\bU=0, \quad \tbU=-\bU. \label{Ito:UAlg}
\end{align}
The first nontrivial equation from the hierarchy based on $U_{1,0}\doteq u$ is the Ito equation: 
\begin{align}\label{Ito:NL}
 u_{x_3x_3}+2(u_{x_1x_1x_3}+3u_{x_1}u_{x_3})_{x_1}=0,
\end{align}
and the transform $u=2(\ln\tau)_{x_1}$ gives us the bilinear form of the Ito equation: 
\begin{align}\label{Ito:BL}
 (D_3^2+2D_1^3D_3)\tau\cdot\tau=0.
\end{align}
Similarly, from the reduction of BKP one can also obtain the dynamical and algebraic relations for the infinite vector $\bu_k$ in the Ito class: 
\bse\label{Ito:ukDyn}
\begin{align}
 &k^6\bu_k=\bLd^6\cdot\bu_k-\frac{1}{2}\bU\cdot(\bO_6\cdot\bLd-\tbLd\cdot\bO_6)\cdot\bu_k, \\
 &\partial_{2j+1}\bu_k=\bLd^{2j+1}\cdot\bU+\bU\cdot\tbLd^{2j+1}-\frac{1}{2}\bU\cdot(\bO_{2j+1}\cdot\bLd-\tbLd\cdot\bO_{2j+1})\cdot\bU, \quad j=0,1,2,\cdots.
\end{align}
\ese
The component $u_k^{(0)}\doteq\phi$ gives us the Lax pair for the Ito equation (the Lax pair for the other members in the hierarchy can be calculated 
in a similar way): 
\bse\label{Ito:Lax}
\begin{align}
 &L^{\mathrm{Ito}}\phi=k^6\phi, \label{Ito:LaxS} \\
 &\phi_{x_3}=(\partial_1^3+3u_{x_1}\partial_1)\phi, \label{Ito:LaxT}
\end{align}
\ese
where the spectral problem is associated with a 6th-order differential operator given by 
\begin{align*}
 L^{\mathrm{Ito}}=\partial_1^6+6u_{x_1}\partial_1^4+9u_{x_1x_1}\partial_1^3+(9u_{x_1x_1x_1}+9u_{x_1}^2+2u_{x_3})\partial_1^2
 +(3u_{x_1x_1x_1x_1}+9u_{x_1}u_{x_1x_1}+u_{x_1x_3})\partial_1.
\end{align*}
One comment here is that in the original paper \cite{Ito80} a Lax pair for Equation \eqref{Ito:NL} was derived from the bilinear B\"acklund 
transform of the equation (cf. (4.11) and (4.12) in \cite{Ito80}) and it is effectively a 4th-order Lax pair having two spectral parameters. 
While from the DL framework the Ito equation arises as the 6-reduction of BKP and it has a 6th-order Lax operator.

\section{Soliton solutions}\label{S:Solution}

Solutions for the soliton equations arising from the DL framework can be obtained naturally. In fact, the $\bU$ defined in \eqref{U} involving 
an integral gives us general solutions for (2+1)-dimensional integrable hierarchies. The reductions of the infinite matrix $\bU$, namely 
the $\bU$ defined in Equations \eqref{A:U}, \eqref{B:U} and \eqref{C:U}, provide solutions to the respective (1+1)-dimensional hierarchies. 
By choosing a specific measure and an integration domain, one then has special classes of solutions for these hierarchies. In this section we only 
consider soliton-type solutions to the hierarchies that arise from our framework. 

\subsection{Solitons for the (2+1)-dimensional hierarchies}\label{S:3DSoliton}
To construct soliton solutions, we introduce a measure containing a finite number of distinct singularities. In this subsection we consider 
solutions for the (2+1)-dimensional hierarchies. 
\paragraph{Solitons for the AKP hierarchy.}
For the AKP hierarchy, one can take a particular measure as 
\begin{align}\label{AKP:Singular}
 \rd\zeta(l,l')=\sum_{i=1}^{N}\sum_{j=1}^{N'}A_{i,j}\delta(l-k_i)\delta(l'-k'_j)\rd l\rd l',
\end{align}
from which one can easily see that now singularities $k_i$ and $k'_j$ are introduced into the measure. This now turns out to be a $\overline{\partial}$ 
problem (cf. \cite{AC91}) and the infinite matrix $\bU$ defined in \eqref{U} and the linear integral equation \eqref{IntEq} can be reformulated as 
\begin{align*}
 \bU=\sum_{i=1}^N\sum_{j=1}^{N'}A_{i,j}\bu_{k_i}\tbc_{k'_j}\sigma_{k'_j}, \quad
 \bu_k+\sum_{i=1}^N\sum_{j=1}^{N'}A_{i,j}\frac{\rho_k\sigma_{k_j'}}{k+k_j'}\bu_{k_i}=\rho_k\bc_k.
\end{align*}
If one now takes $k$ to be $k_i$ in the above relation, it becomes a set of linear equations for the infinite vector $\bu_{k_i}$ and therefore 
the explicit expression of $\bU$ is obtained, in other words, the $(N,N')$-soliton solution for the AKP hierarchy is constructed. In practice, 
we introduce the generalised Cauchy matrix $\bM$ defined as 
\begin{align}\label{AKP:M}
 \bM=(M_{j,i})_{N'\times N}, \quad M_{j,i}=\frac{\rho_{k_i}\sigma_{k_j'}}{k_i+k_j'}, \quad 
 \rho_k=\exp\Big(\sum_{j=1}^{\infty}k^jx_j\Big), \quad \sigma_{k'}=\exp\Big(-\sum_{j=1}^{\infty}(-k')^jx_j\Big),
\end{align}
and $\bA=(A_{i,j})_{N \times N'}$ is an arbitrary matrix, and consequently the entries in the infinite matrix $\bU$ can therefore be expressed 
by\footnote{The formula here follows from the convention in \cite{FN16} while the expression given in \cite{HJN16} is written in a reverse way.}
\begin{align}\label{AKP:NLSoliton}
 U_{i,j}=\br^{\mathrm{T}}\bK^i(1+\bA\bM)^{-1}\bA\bK'^{j}\bs,
\end{align}
where the vectors $\br$, $\bs$ and the matrices $\bK$, $\bK'$ are given by 
\begin{align*}
 \br=(\rho_{k_1},\cdots,\rho_{k_N})^{\mathrm{T}}, \quad \bs=(\sigma_{k'_1},\cdots,\sigma_{k'_{N'}})^{\mathrm{T}}, \quad 
 \bK=\diag(k_1,\cdots,k_N), \quad \bK'=\diag(k_1',\cdots,k_{N'}').
\end{align*}
Similarly one can consider the $\tau$-function defined by $\tau=\det(1+\bOa\cdot\bC)$ together with \eqref{AKP:Singular}, and this gives us the 
explicit formula for the $\tau$-function taking the form 
\begin{align}\label{AKP:BLSoliton}
 \tau=\det(1+\bA\bM),
\end{align}
which solves the corresponding multilinear equations. 

\paragraph{Solitons for the BKP hierarchy.}
In the BKP hierarchy, we take a particular measure as 
\begin{align}\label{BKP:Sigular}
 \rd\zeta(l,l')=\sum_{i,j=1}^{2N}A_{i,j}\delta(l-k_i)\delta(l'-k'_j)\rd l\rd l', \quad A_{i,j}=-A_{j,i}.
\end{align}
The reason why $A_{i,j}$ is antisymmetric is that this treatment preserves the antisymmetry property of the measure in the BKP hierarchy in 
infinite matrix form. After some similar computation one can find the $N$-soliton solution for the BKP hierarchy can be expressed by 
\begin{align}\label{BKP:NLSoliton}
 U_{i,j}=\br^{\mathrm{T}}\bK^i(1+\bA\bM)^{-1}\bA\bK'^{j}\br',
\end{align}
where the generalised Cauchy matrix in this case is given by 
\begin{align}\label{BKP:M}
 \bM=(M_{j,i})_{2N\times 2N}, \quad M_{j,i}=\frac{1}{2}\rho_{k_i}\frac{k_i-k_j'}{k_i+k_j'}\rho_{k_j'}, \quad 
 \rho_k=\exp\Big(\sum_{j=1}^{\infty}k^{2j+1}x_{2j+1}\Big),
\end{align}
and $\br$, $\bs$, $\bK$ and $\bK'$ are defined by 
\begin{align*}
 \br=(\rho_{k_1},\cdots,\rho_{k_{2N}})^{\mathrm{T}}, \quad \br'=(\rho_{k_1'},\cdots,\rho_{k_{2N}'})^{\mathrm{T}}, \quad 
 \bK=\diag(k_1,\cdots,k_{2N}), \quad \bK'=\diag(k_1',\cdots,k_{2N'}'),
\end{align*}
and $\bA=(A_{i,j})_{2N\times 2N}$ is a skew-symmetric matrix. 
The solution to the multilinear forms can be expressed by $\tau$ and following the definition of that function in Subsection \ref{S:BKP}, we have 
\begin{align}\label{BKP:BLSoliton}
 \tau^2=\det(1+\bA\bM).
\end{align}
The $\tau$-function itself in the BKP hierarchy can be expressed by a Pfaffian apparently because $\bA$ and $\bM$ are both skew-symmetric matrices. 

\paragraph{Solitons for the CKP hierarchy.}
Likewise we take the symmetric measure in the CKP hierarchy as follow:
\begin{align}\label{CKP:Sigular}
 \rd\zeta(l,l')=\sum_{i,j=1}^{N}A_{i,j}\delta(l-k_i)\delta(l'-k'_j)\rd l\rd l', \quad A_{i,j}=A_{j,i}.
\end{align}
Similarly to the BKP hierarchy, the reason why we require $A_{i,j}$ symmetric is that this condition can preserve the symmetry condition of the measure 
in the CKP hierarchy. We then define $\br$, $\br'$ as 
\begin{align*}
 \br=(\rho_{k_1},\cdots,\rho_{k_{N}})^{\mathrm{T}}, \quad \br'=(\rho_{k_1'},\cdots,\rho_{k_N'})^{\mathrm{T}},
\end{align*}
and let $\bK$ and $\bK'$ be exactly the same as those in the AKP hierarchy. Suppose $\bA=(A_{i,j})_{N\times N}$ is a symmetric matrix, 
and introducing the generalised Cauchy matrix 
\begin{align}\label{CKP:M}
 \bM=(M_{j,i})_{N\times N}, \quad M_{j,i}=\frac{\rho_{k_i}\rho_{k_j'}}{k_i+k_j'}, \quad \rho_k=\exp\Big(\sum_{j=0}^{\infty}k^{2j+1}x_{2j+1}\Big),
\end{align}
we have the $N$-soliton solution for the nonlinear form of the CKP hierarchy and it is determined by the expression for the entry $U_{i,j}$ as follow: 
\begin{align}\label{CKP:NLSoliton}
 U_{i,j}=\br^{\mathrm{T}}\bK^i(1+\bA\bM)^{-1}\bA\bK'^{j}\br'.
\end{align}
Similarly, the $\tau$-function takes the form 
\begin{align}\label{CKP:BLSoliton}
 \tau=\det(1+\bA\bM),
\end{align}
providing the $N$-soliton solution to the multilinear form of the CKP hierarchy.

\subsection{Solitons for the (1+1)-dimensional hierarchies}\label{S:2DSoliton}

\paragraph{Solitons for the reductions of AKP.} For the $N$-reduction of the AKP hierarchy we already have the general expressions for the 
infinite matrix $\bU$, namely \eqref{A:U} and the the reduced integral equation \eqref{A:IntEq}. One can restrict the measures $\ld_j(l)$ to be a 
particular form involving singularities as 
\begin{align}\label{A:Sigular}
 \rd\ld_{j}(l)=\sum_{j'=1}^{N_j}A_{j,j'}\delta(l-k_{j,j'})\rd l,
\end{align}
and as a result the infinite matrix $\bU$ and the linear integral equation \eqref{A:IntEq} under the particular measure can be written as 
\begin{align*}
 \bu_k+\sum_{j=1}^N\sum_{j'=1}^{N_j}A_{j,j'}\frac{\rho_k\sigma_{-\oa^jk_{j,j'}}}{k-\oa^jk_{j,j'}}\bu_{k_{j,j'}}=\rho_k\bc_k, \quad 
 \bU=\sum_{j=1}^N\sum_{j'=1}^{N_j}A_{j,j'}\bu_{k_{j,j'}}\tbc_{-\oa^jk_{j,j'}}\sigma_{-\oa^jk_{j,j'}}.
\end{align*}
Here $N$ denotes the $N$-reduction, namely $\oa^N=1$, while $N_j$ denotes the number of solitons in solution. So are the notations for the reduced 
hierarchy of BKP and CKP. Taking $k=k_{i,i'}$ in the above relations, one can obtain the expression for $\bU$ by solving $\bu_{k_{i,i'}}$ in the 
first equation given above. The expression of the entries in $\bU$ is 
\begin{align}\label{A:NLSoliton}
 U_{i,j}=\br^{\mathrm{T}}\bK^i(1+\bA\bM)^{-1}\bA\bK'^{j}\bs,
\end{align}
in which the generalised Cauchy matrix is defined by 
\begin{align}\label{A:M}
 \bM=(M_{(j,j'),(i,i')})_{j,i=1,\cdots,N,j'=1,\cdots,N_j,i'=1,\cdots,N_i}, \quad 
 M_{(j,j'),(i,i')}=\frac{\rho_{k_{i,i'}}\sigma_{-\oa^jk_{j,j'}}}{k_{i,i'}-\oa^jk_{j,j'}}
\end{align}
with $\rho_k$ and $\sigma_k$ defined as those in \eqref{AKP:M}. The matrix $\bM$ should be understood as a $N\times N$ block matrix in which 
the $(j,i)$-entry is a rectangular matrix of size $N_j\times N_i$. In other words, the indices $(j,i)$ denote the blocks and the indices 
$(j',i')$ denote the entries in each block. While the vectors $\br$ and $\bs$, and the matrices $\bK$, $\bK'$ and $\bA$ are given by 
\begin{align*}
 &\br=(\rho_{k_{1,1}},\cdots,\rho_{k_{1,N_1}};\cdots;\rho_{k_{j,1}},\cdots,\rho_{k_{j,N_j}};\cdots;\rho_{k_{N,1}},\cdots,\rho_{k_{N,N_N}})^{\mathrm{T}}, \\
 &\bs=(\sigma_{-\oa k_{1,1}},\cdots,\sigma_{-\oa k_{1,N_1}};\cdots;\sigma_{-\oa^j k_{j,1}},\cdots,\sigma_{-\oa^j k_{j,N_j}};\cdots;
 \sigma_{-\oa^N k_{N,1}},\cdots,\sigma_{-\oa^N k_{N,N_N}})^{\mathrm{T}}, \\
 &\bK=\diag(k_{1,1},\cdots,k_{1,N_1};\cdots;k_{j,1},\cdots,k_{j,N_j};\cdots;k_{N,1},\cdots,k_{N,N_N}), \\
 &\bK'=\diag(-\oa k_{1,1},\cdots,-\oa k_{1,N_1};\cdots;-\oa^j k_{j,1},\cdots,-\oa^j k_{j,N_j};\cdots;-\oa^N k_{N,1},\cdots,-\oa^N k_{N,N_N}), \\
 &\bA=\diag(A_{1,1},\cdots,A_{1,N_1};\cdots;A_{j,1},\cdots,A_{j,N_j};\cdots;A_{N,1},\cdots,A_{N,N_N}).
\end{align*}
The $\tau$-function takes the form 
\begin{align}\label{A:BLSoliton}
 \tau=\det(1+\bA\bM)
\end{align}
with $\bM$ defined as \eqref{A:M}. 

The formulae for $U_{i,j}$ and $\tau$ govern the soliton solutions for the equations arising as the reductions of the AKP hierarchy. 
For instance, the cases $N=2,3,4$ are corresponding to the solitons of the KdV, BSQ and gHS hierarchies, respectively.

\paragraph{Solitons for the reductions of BKP.} 
One can take the particular measures $\ld_j(l)$ like \eqref{A:Sigular}, and as a result the infinite matrix $\bU$ and the linear integral 
equation \eqref{B:IntEq} under the particular measures can be written as 
\begin{align*}
 &\bu_k+\sum_{j=1}^N\sum_{j'=1}^{N_j}A_{j,j'}\frac{1}{2}\rho_k\frac{k+\oa^jk_{j,j'}}{k-\oa^jk_{j,j'}}\rho_{-\oa^jk_{j,j'}}\bu_{k_{j,j'}}
 -\sum_{j=1}^N\sum_{j'=1}^{N_j}A_{j,j'}\frac{1}{2}\rho_k\frac{k-k_{j,j'}}{k+k_{j,j'}}\rho_{k_{j,j'}}\bu_{-\oa^jk_{j,j'}}=\rho_k\bc_k, \\
 &\bU=\sum_{j=1}^N\sum_{j'=1}^{N_j}A_{j,j'}\bu_{k_{j,j'}}\tbc_{-\oa^jk_{j,j'}}\rho_{-\oa^jk_{j,j'}}
 -\sum_{j=1}^N\sum_{j'=1}^{N_j}A_{j,j'}\bu_{-\oa^jk_{j,j'}}\tbc_{k_{j,j'}}\rho_{k_{j,j'}}.
\end{align*}
Similarly we can calculate the expressions of the entries in $\bU$ and obtain: 
\begin{align}\label{B:NLSoliton}
 U_{i,j}=
 \left(
 \begin{array}{c}
  \br \\
  \br'
 \end{array}
 \right)^{\mathrm{T}}
 \left(
 \begin{array}{cc}
  \bK & 0 \\
  0 & \bK'
 \end{array}
 \right)^i
 \left(1+
 \left(
 \begin{array}{cc}
  \bA & 0 \\
  0 & -\bA
 \end{array}
 \right)
 \left(
 \begin{array}{cc}
  \bM & 0 \\
  0 & \bM'
 \end{array}
 \right)
 \right)^{-1}
 \left(
 \begin{array}{cc}
  \bA & 0 \\
  0 & -\bA
 \end{array}
 \right)
 \left(
 \begin{array}{cc}
  \bK' & 0 \\
  0 & \bK
 \end{array}
 \right)^j
 \left(
 \begin{array}{c}
  \br' \\
  \br
 \end{array}
 \right),
\end{align}
in which the generalised Cauchy matrix is a block matrix defined by 
\bse\label{B:M}
\begin{align}
 &\bM=(M_{(j,j'),(i,i')})_{j,i=1,\cdots,N,j'=1,\cdots,N_j,i'=1,\cdots,N_i}, \quad 
 M_{(j,j'),(i,i')}=\frac{1}{2}\rho_{k_{i,i'}}\frac{k_{i,i'}+\oa^jk_{j,j'}}{k_{i,i'}-\oa^jk_{j,j'}}\rho_{-\oa^jk_{j,j'}}, \\
 &\bM'=(M_{(j,j'),(i,i')}')_{j,i=1,\cdots,N,j'=1,\cdots,N_j,i'=1,\cdots,N_i}, \quad 
 M_{(j,j'),(i,i')}'=\frac{1}{2}\rho_{-\oa^ik_{i,i'}}\frac{-\oa^ik_{i,i'}-k_{j,j'}}{-\oa^ik_{i,i'}+k_{j,j'}}\rho_{k_{j,j'}},
\end{align}
\ese
where $\rho_k$ is given in \eqref{BKP:M}, and the vectors $\br$ and $\br'$, and the matrices $\bK$, $\bK'$ and $\bA$ are given by 
\begin{align*}
 &\br=(\rho_{k_{1,1}},\cdots,\rho_{k_{1,N_1}};\cdots;\rho_{k_{j,1}},\cdots,\rho_{k_{j,N_j}};\cdots;\rho_{k_{N,1}},\cdots,\rho_{k_{N,N_N}})^{\mathrm{T}}, \\
 &\br'=(\rho_{-\oa k_{1,1}},\cdots,\rho_{-\oa k_{1,N_1}};\cdots;\rho_{-\oa^jk_{j,1}},\cdots,\rho_{-\oa^jk_{j,N_j}};\cdots;
 \rho_{-\oa^Nk_{N,1}},\cdots,\rho_{-\oa^Nk_{N,N_N}})^{\mathrm{T}}, \\
 &\bK=\diag(k_{1,1},\cdots,k_{1,N_1};\cdots;k_{j,1},\cdots,k_{j,N_j};\cdots;k_{N,1},\cdots,k_{N,N_N}), \\
 &\bK'=\diag(-\oa k_{1,1},\cdots,-\oa k_{1,N_1};\cdots;-\oa^j k_{j,1},\cdots,-\oa^j k_{j,N_j};\cdots;-\oa^N k_{N,1},\cdots,-\oa^N k_{N,N_N}), \\
 &\bA=\diag(A_{1,1},\cdots,A_{1,N_1};\cdots;A_{j,1},\cdots,A_{j,N_j};\cdots;A_{N,1},\cdots,A_{N,N_N}).
\end{align*}
The $\tau$-function is determined by 
\begin{align}\label{B:BLSoliton}
 \tau^2=\det\Bigg[1+
 \left(
 \begin{array}{cc}
  \bA & 0 \\
  0 & -\bA
 \end{array}
 \right)
 \left(
 \begin{array}{cc}
  \bM & 0 \\
  0 & \bM'
 \end{array}
 \right)
 \Bigg],
\end{align}
which solves the multilinear forms of the reduced hierarchies from BKP. 

Following the above general formulae for $U_{i,j}$ and $\tau$, we can see that the particular case when $N=3$ gives us the solitons solutions 
to \eqref{SK:NL} and \eqref{SK:BL}, and the case when $N=6$ gives us the soliton solutions to \eqref{Ito:NL} and \eqref{Ito:BL}.

\paragraph{Solitons for the reductions of CKP.}
Likewise we impose \eqref{A:Sigular} on the reduced integral equation \eqref{C:IntEq} and the reduced infinite matrix $\bU$ \eqref{C:U} in the CKP hierarchy. 
This gives us the following relations: 
\begin{align*}
 &\bu_k+\sum_{j=1}^N\sum_{j'=1}^{N_j}A_{j,j'}\frac{\rho_k\rho_{-\oa^jk_{j,j'}}}{k-\oa^jk_{j,j'}}\bu_{k_{j,j'}}
 +\sum_{j=1}^N\sum_{j'=1}^{N_j}A_{j,j'}\frac{\rho_k\rho_{k_{j,j'}}}{k+k_{j,j'}}\bu_{-\oa^jk_{j,j'}}=\rho_k\bc_k, \\
 &\bU=\sum_{j=1}^N\sum_{j'=1}^{N_j}A_{j,j'}\bu_{k_{j,j'}}\tbc_{-\oa^jk_{j,j'}}\rho_{-\oa^jk_{j,j'}}
 +\sum_{j=1}^N\sum_{j'=1}^{N_j}A_{j,j'}\bu_{-\oa^jk_{j,j'}}\tbc_{k_{j,j'}}\rho_{k_{j,j'}}.
\end{align*}
If one introduces the following block generalised Cauchy matrices:
\bse\label{C:M}
\begin{align}
 &\bM=(M_{(j,j'),(i,i')})_{j,i=1,\cdots,N,j'=1,\cdots,N_j,i'=1,\cdots,N_i}, \quad 
 M_{(j,j'),(i,i')}=\frac{\rho_{k_{i,i'}}\rho_{-\oa^jk_{j,j'}}}{k_{i,i'}-\oa^jk_{j,j'}}, \\
 &\bM'=(M_{(j,j'),(i,i')})_{j,i=1,\cdots,N,j'=1,\cdots,N_j,i'=1,\cdots,N_i}, \quad 
 M_{(j,j'),(i,i')}'=\frac{\rho_{-\oa^ik_{i,i'}}\rho_{k_{j,j'}}}{-\oa^ik_{i,i'}+k_{j,j'}},
\end{align}
\ese
the entries $U_{i,j}$ in the infinite matrix $\bU$ in this class can be written as 
\begin{align}\label{C:NLSoliton}
 U_{i,j}=
 \left(
 \begin{array}{c}
  \br \\
  \br'
 \end{array}
 \right)^{\mathrm{T}}
 \left(
 \begin{array}{cc}
  \bK & 0 \\
  0 & \bK'
 \end{array}
 \right)^i
 \left(1+
 \left(
 \begin{array}{cc}
  \bA & 0 \\
  0 & \bA
 \end{array}
 \right)
 \left(
 \begin{array}{cc}
  \bM & 0 \\
  0 & \bM'
 \end{array}
 \right)
 \right)^{-1}
 \left(
 \begin{array}{cc}
  \bA & 0 \\
  0 & \bA
 \end{array}
 \right)
 \left(
 \begin{array}{cc}
  \bK' & 0 \\
  0 & \bK
 \end{array}
 \right)^j
 \left(
 \begin{array}{c}
  \br' \\
  \br
 \end{array}
 \right),
\end{align}
where $\br$, $\br'$, $\bK$ and $\bK'$ and $\bA$ are the same as those given in the soliton solutions for the reduced hierarchies from BKP. 
The $\tau$-function given by 
\begin{align}\label{C:BLSoliton}
 \tau=\det\Bigg[1+
 \left(
 \begin{array}{cc}
  \bA & 0 \\
  0 & \bA
 \end{array}
 \right)
 \left(
 \begin{array}{cc}
  \bM & 0 \\
  0 & \bM'
 \end{array}
 \right)
 \Bigg]
\end{align}
provides soliton solutions to the hierarchies in the multilinear form. 

In this case, the expressions of the $U_{i,j}$ and the $\tau$-function provide soliton solutions for the $N$-reduced hierarchies from CKP. 
For example, the $N=3$ case comprises the solitons for the equations in the SK and KK family, and the $N=4$ case provides those for 
equations in the HS family.

\section{Conclusions}\label{S:Concl}

We presented a unified framework to understand (2+1)- and (1+1)-dimensional soliton hierarchies associated with scalar linear integral equations. 
The framework provides many of the requisite integrability characteristics such as the solution structure (including explicit solutions such as 
soliton solutions), the associated linear problem (Lax pair), Miura-type transforms among different nonlinear forms and the multilinear form for 
the $\tau$-function for each soliton hierarchy. All well-known soliton hierarchies associated with a scalar differential spectral problem are covered. 

We obtained all the (1+1)-dimensional soliton hierarchies from reductions of (2+1)-dimensional models by imposing certain conditions on the measures 
and integration domains in the corresponding linear integral equations. As a by-product, we attained a richer solution structure for the higher-rank soliton 
equations where block Cauchy matrices are involved. Furthermore, our approach has some advantage over the ``standard approach'' to the KP hierarchy which 
exploits pseudo-differential operators in that in the DL framework it is not necessary to single out a particular flow variable (usually called $x\doteq x_1$) 
in order to set up the framework. 

Other integrability characteristics such as recursion operators and Hamiltonian structures were not considered in the paper. Nevertheless from the view 
point of the DL, they can be obtained by considering squared eigenfunctions, cf. the procedure given in \cite{CWN86}. However, the integrability of 
the equations in the hierarchies that we obtained is self-evident from the fact that they possess infinite families of explicit solutions from our framework. 

While, the starting point in the current paper is the DL scheme for the KP hierarchy associated with the $A_{\infty}$ algebra.  The other examples such as 
BKP and CKP are the hierarchies associated with the sub-algebras $B_\infty$ and $C_\infty$ which are contained in the AKP case. And so are the reduced 
(1+1)-dimensional hierarchies arising from the AKP, BKP and CKP hierarchies. There also exist (3+1)-dimensional soliton equations beyond this case. 
A recent result \cite{JN14} shows that the KP hierarchy has an elliptic extension and the model also leads to elliptic extensions of the ``sub-hierarchies'' 
(e.g. the elliptic KdV hierarchy \cite{NP03}). However, the algebras hidden behind these models are not yet clear. 

There also exists the so-called DKP hierarchy, named after the infinite-dimensional algebra $D_\infty$. 
However, the DKP hierarchy is a sub-case of the two-component AKP hierarchy, 
whose corresponding linear structure is beyond the integral equation \eqref{IntEq}. 
We will report the relevant results elsewhere in the future.

\subsection*{Acknowledgements}
WF benefited a lot from discussions with Allan Fordy, and he was supported by a Leeds International Research Scholarship (LIRS) as well as 
a small grant from the School of Mathematics. FWN was partially supported by EPSRC (Ref. EP/I038683/1).

\end{document}